# Broadcast Cooperation Strategies for Two Colocated Users

Avi Steiner*,Amichai Sanderovich* and Shlomo Shamai (Shitz)*

*Technion—Israel Institute of Technology

Haifa 32000, Israel

October 18, 2018

**Abstract**

This work considers the problem of communication from a single transmitter, over a network with colocated users, through an independent block Rayleigh fading channel. The colocation nature of the users allows cooperation, which increases the overall achievable rate, from the transmitter to the destined user. The transmitter is ignorant of the fading coefficients, while receivers have access to perfect channel state information (CSI). This gives rise to the broadcast approach used by the transmitter. The broadcast approach facilitates reliable transmission rates adapted to the actual channel conditions, designed to maximize average throughput. It also allows in our network setting to improve the cooperation between the colocated users. With the broadcast approach, users can decode something out of the message, with almost any fading realization. The better the channel quality, the more layers that can be decoded. Such an approach is useful when considering average rates, rather than outage vs. rate. The cooperation between the users is performed over an additive white Gaussian channels (AWGN), with a relaying power constraint, and unlimited bandwidth. One type of cooperation studied is the amplify-forward (AF) cooperation. Another is the Wyner-Ziv (WZ) compression and forwarding (CF) technique. And finally, decode and forward (DF) cooperation is investigated. In this paper, we extend these methods using the broadcast approach, for the case of relaxed decoding delay constraint. For this case a separated processing of the layers, which includes multi-session cooperation is shown to be beneficial. Further, closed form expressions for infinitely many AF sessions and recursive expressions for the more complex WZ are given. Numerical results for the various cooperation strategies demonstrate the efficiency of multi-session cooperation. Our results can be extended straightforwardly to a setting of a single transmitter sending common information for two cooperating users.

# I. INTRODUCTION

In recent years, interest in communication networks has increased, and various applications of it, such as sensor networks [1],[2],[3] energy sensitive networks [4],[5] and Ad-hoc networks [6], have gained popularity. In this field, networks with colocated receivers and colocated transmitters constitute a substantial part, since it allows increased cooperation [7], thus improving the overall network throughput [8], [9]. Specifically, many works deal with the various aspects of such cooperation, that is, transmitters cooperation [10],[11], receivers cooperation [12],[13],[14] [15] and [16] and both receivers and transmitters cooperation [17],[18]. In source related networks, such as the sensors network, the cooperation is slightly different, since the objective is to convey a source with a distortion (e.g. the reach-back problem [19]), rather than ensuring reliable communication. The compress and forward (CF) and amplify and forward (AF) techniques make use of lossy source coding techniques, to ensure high communication rates, when the cooperative receiver does not decode the message. This is done in [20],[21], among many others. Here, we deal with one transmitter that sends the same information to two colocated users, through independent, block Rayleigh fading channels [22]. Such channels have zero Shannon capacity, and usually one turns to rate versus outage probability [23]. When considering the average throughput or delay as figures of merit, it is beneficial to use the broadcast approach [24]. The broadcast strategy for a single-user facilitates reliable transmission rates adapted to the actual channel conditions, without providing any feedback from the receiver to the transmitter [24], [25]. The single-user broadcasting approach hinges on the broadcast channel, which was first explored by Cover [26]. In a broadcast channel, a single transmission is directed to a number of receivers, each enjoying possibly different channel conditions, reflected in their received signal to noise ratio (SNR). Here, every fading gain is associated with another user, thus there can be no outage. The higher the fading gain, the higher is the achievable rate.

The broadcast approach has been used in [27] for a two hop relay channel, where the efficiency of ad-hoc cooperation in a two-hop relay setting was demonstrated, when a direct link from source to destination is not available. Several broadcasting strategies were designed for relaying techniques such as DF, AF, and CF. In our setting, a direct link from source to destination exists in addition to the cooperation link, which motivates multi-session cooperation, and different broadcasting approaches for maximizing average throughput. In [15], a cooperation among densely packed $K$ colocated receivers is studied, where the users with better channel conditions decode the message faster, and join the transmission to the destined user, thus allowing the destined user to decode the original message even if a severe fading occurs on the source destination link. Notice that transmission and cooperation in [15] take place within a single block, whereas in our work we consider multi-session cooperation which starts after the transmission of the previous block was complete. In [16], a similar network setting is considered, with a single source transmitting to two colocated users, where a Wyner-Ziv (WZ) CF single session cooperation is studied. The WZ-CF in [16] does not assume knowledge of the actual fading realization on the source-destination link. In our work, we assume that prior to the WZ compression the destination sends the relay its actual fading gain, and thus we incorporate continuous broadcasting with optimal power allocation, as the transmitter views a single equivalent fading gain. In which case cooperation schemes, such as AF or CF, can be treated as an equivalent fading channel (usually non-Rayleigh) between the transmitter and the destination, and an adapted broadcast approach can be used.

In this paper, we consider the case where the two receivers can cooperate between themselves, so that they can improve reception at destination receiver, via DF or via source related techniques such as AF or CF. Since these users are colocated, the probability of a multi-path non-line-of-sight channel, such as the channel from the transmitter, is low, so the cooperation takes place over the additive white Gaussian noise (AWGN) channel, with a relaying power constraint, and unlimited bandwidth. In addition to single session cooperation, we study multi-session cooperation schemes, like was done by [28] for the binary erasure channel. By combining the broadcasting approach with multi session cooperation, the efficiency of each session is increased by reducing information layers that were decoded in previous sessions. This way, we can surpass the naive cooperation performance.

The rest of the paper is organized as follows. The main contributions of this work are described in section II, and the channel model is specified in III. Upper and lower bounds are stated in section IV. Section V deals with cooperation through the simpler amplify and forward, and section VI improves the achievable rates of the previous section, by using Wyner-Ziv (CF). Section VII describes the broadcast approach with DF cooperation. Then, section VIII gives numerical results, comparing the achievable rates via the various cooperation schemes. Finally, the paper ends with concluding remarks, and a discussion of a straightforward extension of our results to the case of receiving common information at two cooperating colocated end-users.

## II. Main Contributions

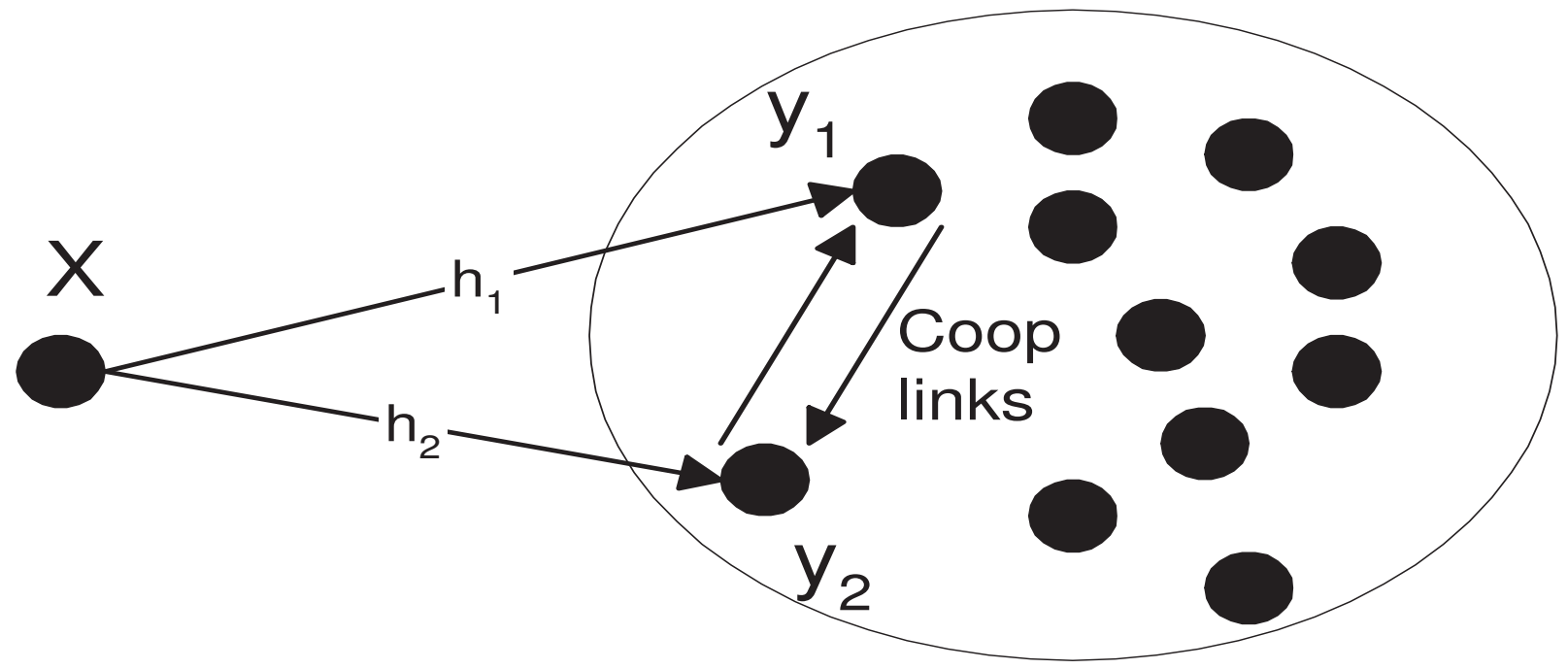

Fig. 1. Schematic diagram of a source transmitting to two colocated users, with multi-session cooperation.

We study the problem of single-user broadcasting [24], [25] over fading channels of a wireless network with colocated cooperating users. The transmitter does not posses any knowledge of channel state information (CSI), whereas the receivers have access to perfect CSI. The wireless network setting is illustrated in Figure 1.

Three types of cooperation strategies are considered (AF, CF, and DF). The first is based on the low complexity amplify-and-forward (AF) relaying by a network user to the destination user, over the cooperation link. The simplest form of AF cooperation is the naive AF, where the relaying user simply scales its input and forwards it to the destined user, who jointly decodes the signal from the direct link, and the relay. In this case an equivalent fading gain may be formulated between transmitter and destination receiver, and maximal achievable rate is derived in closed form. A more efficient form of single session AF is the separate preprocessing, where the colocated users exchange the values of the estimated fading gains, then individually decode the layers up to the smallest fading gain. The relaying user subtracts this decoded common information from its received signal and performs AF to the destined user. Achievable rates are computed for this case using sub-optimal power distribution for the broadcast approach at the source. An extension of this approach is when the two colocated users perform this process repeatedly. This form of cooperation is referred to as multi-session cooperation, where repeatedly separate preprocessing is followed by a transmission of cooperation information at both relay and destination sides. The preprocessing basically includes individual decoding with the available received information from the direct link and previous cooperation sessions. During the cooperation sessions the transmission of the next block already takes place. This means that our multi-session cooperation introduces additional decoding delays, without reducing the overall throughput. On the contrary, multi-session cooperation allows increasing the overall throughput by enhancing the decoding capability of an individual user, thus requiring less retransmissions. However, the simultaneous transmission of the next block requires that overall, in each block time slot, processing of many blocks be performed, as well as cooperation channel uses. Thus requiring multiple parallel cooperation channels between the cooperating users. Hence the cooperation link has unlimited bandwidth, as illustrated in Figure 3. In order to incorporate practical constraints on the multi-session approach, the total power of a multi-session cooperation is restricted to that of a single session cooperation. A fixed power allocation $P_r$ per cooperation block is assumed, thus for multi-session the power is split between sessions such that the overall cooperation transmission power does not exceed $P_r$.

The capacity of such channel is $C_{coop} = P_r$ (4). Since the two latter schemes of AF, that is the naive AF and the separate preprocessing AF, can not efficiently use the unlimited bandwidth, a narrow-band cooperation channel is used for these two schemes, with $C_{coop} = \log(1 + P_r)$.

The second cooperation strategy is based on the Wyner-Ziv [29] CF strategy. The simplest form of CF cooperation here is the narrow-band naive CF, where the relaying user performs WZ-CF over a link of capacity $C_{coop} = \log(1+P_r)$. Prior to the WZ compression, the destination informs the relay of the actual fading gain it has estimated. Using this information, the transmitter adapts the broadcast approach to be optimized for the equivalent fading channel the destination views, while assuming the destination performs optimal decoding using its own copy of the signal from the direct link and the WZ compressed signal forwarded over the cooperation link. Like in naive AF, an equivalent fading gain may be formulated between transmitter and destination receiver, of an associated point-to-point equivalent channel, and maximal achievable rate is derived in closed form. The extension of the WZ-CF approach to the unlimited bandwidth power limited cooperation link is straightforward. A more advanced form of cooperation is the multi-session CF, which is performed adhering to successive refinement WZ. Recursive expressions for the equivalent fading gain are derived. Numerical results here show the high efficiency of CF, which highly approximates the joint decoding upper bound, already in a narrow-band naive

CF cooperation, and clearly outperforms DF cooperation.

We consider also DF cooperation, where the cooperating users are colocated [30], and the source transmitter performs broadcasting. However, the DF cooperation is not suitable for multi-session cooperation. That is, after a first session of cooperation, where the relay has sent to the destination its decoded layers (on top of those decoded independently at the destination), the destination cannot send any information back to the relay in order to decode more layers. In DF the achievable rate is limited by the maximal rate achievable independently by each user. This type of limitation does not exist in AF and CF cooperation. As already known, such cooperation is beneficial to both users, with the stronger and weaker channels [31]. Similarly to single-session CF, a wide-band DF cooperation can also be used here, and numerical results show that wide-band DF cooperation closely approximates the DF upper bound.

Our results can be straightforwardly extended to the case of a single transmitter sending common information to two cooperating users. See section IX for more details.

## III. Channel Model

Consider the following single-input multiple-output (SIMO) channel (we use boldfaced letters for vectors) ,

$$\mathbf{y}_i = h_i \mathbf{x}_s + \mathbf{n}_i \quad , \; i = 1, 2 \tag{1}$$

where $\mathbf{y}_i$ is a received vector by user $i$, of length $L$, which is also the transmission block length. $\mathbf{x}_s$ is the original source transmitted vector. $\mathbf{n}_i$ is the additive noise vector, with elements that are complex Gaussian i.i.d with zero mean and unit variance, denoted $\mathcal{CN}(0, 1)$, and $h_i$ is the (scalar) fading coefficient. The fading $h_i$ is assumed to be perfectly known by the receivers. The fading $h_i$ is distributed according to the Rayleigh distribution $h_i \sim \mathcal{CN}(0, 1)$, and remains constant for the duration of the transmission (block fading). This also means that the two users have equal average SNR, which is realistic due to the colocation assumption. The source transmitter has no CSI, and the power constraint at the source is given by $E|x_s|^2 \leq P_s$. $E$ stands for the expectation operator. Without loss of generality we assume here that the destination is user $i = 1$. The cooperation channels between the users are modelled by AWGN channels as follows

$$\begin{aligned} \mathbf{y}_{2,1}^{(k)} &= \mathbf{x}_1^{(k)} + \mathbf{w}_1^{(k)} \\ \mathbf{y}_{1,2}^{(k)} &= \mathbf{x}_2^{(k)} + \mathbf{w}_2^{(k)} \end{aligned} \tag{2}$$

where $\mathbf{y}_{2,1}^{(k)}$ is the second user's received cooperation vector (of length $L$) from the destination ($i = 1$), on the $k^{th}$ cooperation link, and vise-versa for $\mathbf{y}_{1,2}^{(k)}$. $\mathbf{x}_i^{(k)}$ is the cooperation signal from user $i$, on the $k^{th}$ cooperation link, and $\mathbf{w}_i$ is the noise vector with i.i.d elements distributed according to $\mathcal{CN}(0, 1)$. For a single session cooperation $k = 1$, and the power of $x_i^{(1)}$ is limited by $E|x_i^{(1)}|^2 \leq P_r$ (for $i = 1, 2$). However, for a wide-band cooperation $k = 1, 2, ..., K$, which models $K-$parallel cooperation channels for each user. The power constraint here is specified by $E \sum\limits_{k=1}^{K} |x_i^{(k)}|^2 \leq P_r$ (for $i = 1, 2$). So $K$ is the bandwidth expansion that results from the multi-session cooperation.

Naturally, the link capacity of a single session narrow-band cooperation is given by

$$C_{coop,NB} = \log(1 + P_r). \tag{3}$$

In the limit of $K \to \infty$ with a power constraint for multi-session cooperation, the cooperation link capacity is given by

$$C_{coop,WB} = \int\limits_0^\infty dR(s) = \int\limits_0^\infty \rho(s)ds = P_r, \tag{4}$$

where $dR(s)$ is the fractional rate in session associated with parameter $s$, and $dR(s) = \log(1 + \rho(s)ds)$. The fractional power at the $s^{th}$ session is $\rho(s)$. The multi-session power constraint implies $\int\limits_0^\infty \rho(s)ds = P_r$, which justifies the last equality in (4).

In view of a single-session cooperation, the AF strategy cannot use more than the original signal bandwidth in a single session. However, both naive CF and DF approaches may utilize a cooperation channel bandwidth expansion of the form $C_{coop,WB}$ (4) for improving the cooperation efficiency. This is also considered in the following.

## IV. Upper and Lower bounds

In order to evaluate the benefit of cooperation among receivers in a fading channel following the model described in (1)-(2), we bring here some upper and lower bounds.

### A. Lower bounds

One immediate lower bound is the single receiver lower bound. That is, the outage and broadcasting average rates [25] are computed for a single user, assuming there are no available users for cooperation. The distribution of the fading gain of a single user over a Rayleigh channel is given by $F(u) = 1 - e^{-u}$.

*1) Outage lower bound:* The achievable single-level coding average rate is given by

$$R_{outage,LB} = \max_{u_{th}>0} \{(1 - F(u_{th})) \log(1 + u_{th} P_s)\} \tag{5}$$

where the optimal threshold $u_{th}$ which maximizes (5) is given by $u_{th,opt} = \frac{P_s - W(P_s)}{W(P_s)P_s}$. The function $W(x)$ is the Lambert-W function, also known as the omega function.

*2) Broadcasting lower bound:* The average achievable broadcasting rate is given by [25],

$$R_{bs,LB} = e^{-1} - e^{-s_0} + 2E_1(s_0) - 2E_1(1) \tag{6}$$

where $s_0 = 2/(1 + \sqrt{1 + 4P_s})$, and $E_1(x)$ is the exponential integral function $E_1(x) = \int_x^{\infty} dt \frac{e^{-t}}{t}$ for $x \geq 0$.

## B. Upper bound

A natural upper bound here is the joint decoding upper bound. In this case a single receiver with two antennas and optimal processing is assumed. The distribution of an equivalent fading gain of a channel with two fully-cooperating agents is $F_{UB}(u) = 1 - e^{-u} - ue^{-u}$.

*1) Outage upper bound:* An outage bound for fully cooperating users is derived similarly to (5), with $F_{UB}(u)$ as the fading gain distribution function.

*2) Broadcasting upper bound:* The corresponding average broadcasting rate is

$$R_{bs,UB} = \quad s_1 e^{-s_1} - e^{-s_1} - 3E_1(s_1) - (s_0 e^{-s_0} - e^{-s_0} - 3E_1(s_0)) \tag{7}$$

where $s_0$ and $s_1$ are determined by the boundary conditions $I_{UB}(s_0) = P_s$ and $I_{UB}(s_1) = 0$, respectively. The residual interference $I_{UB}(x)$ is given by $I_{UB}(x) = (1 + x - x^2)/x^3$.

*3) Cut-set upper bound:* Another upper bound considered is the classical cut-set bound of the relay channel [32]. Using the relay channel definitions in (1)-(2), and assuming without loss of generality that the destination is user $i = 1$, and only single cooperation session $K = 1$, the cut-set bound for a Rayleigh fading channel is given by:

$$\begin{aligned} C_{cut-set} &= \sup_{p(x_s),p(x_2)} \min\{I(x_s; y_1|h_1) + I(x_2; y_{1,2}), I(x_s; y_1, y_2|h_1, h_2)\} \\ &= \min\{C_{erg}(1) + \log(1 + P_r),\ C_{erg}(2)\} \end{aligned} \tag{8}$$

where the ergodic capacity $C_{erg}(m)$ is the ergodic capacity between a source and $m$ users performing optimal cooperation, which is given by

$$C_{erg}(m) = \int_0^{\infty} u^{m-1} e^{-u} \log(1 + P_s u) du, \quad m = 1, 2, ... \tag{9}$$

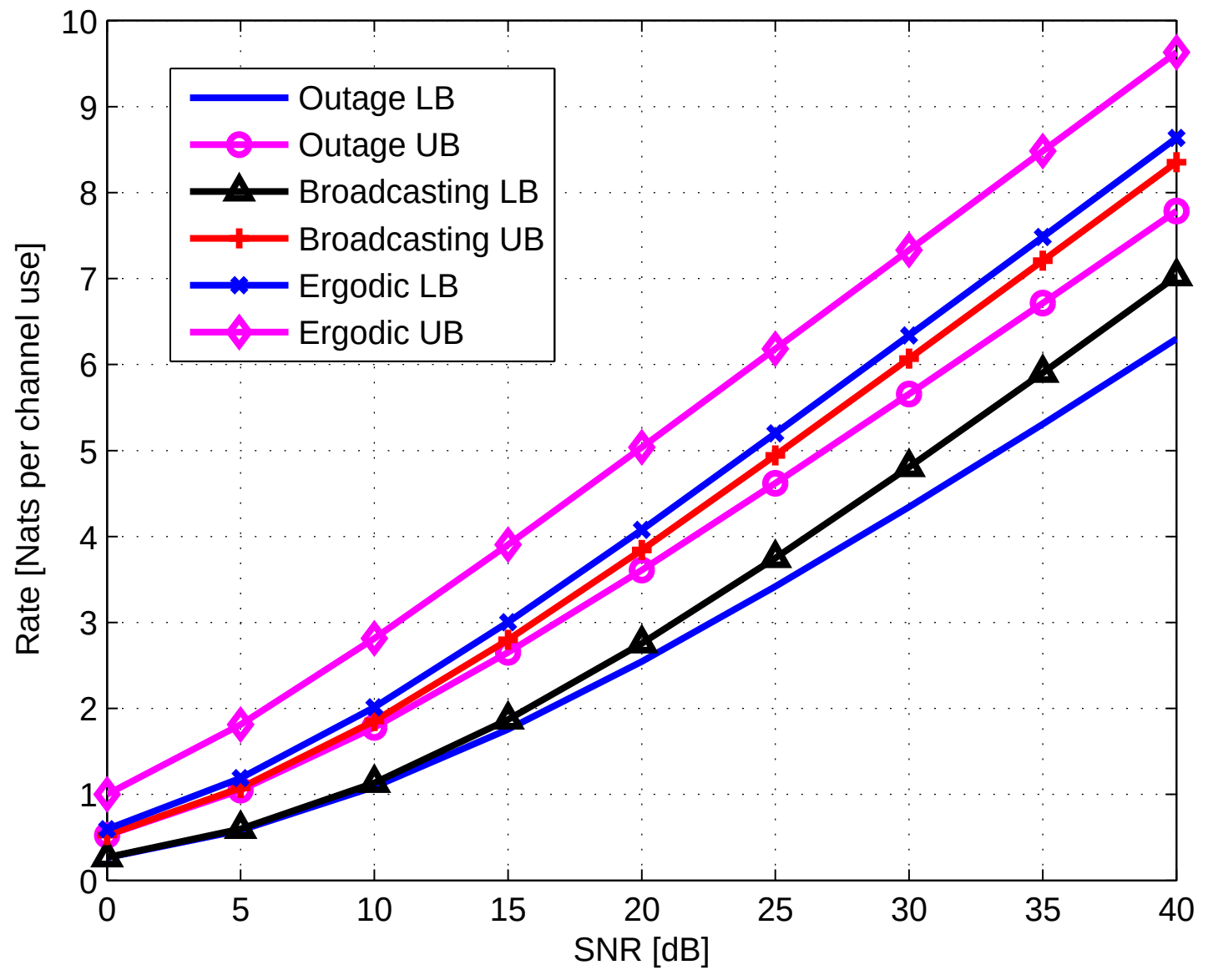

Fig. 2. Outage and broadcasting bounds. 'LB' denotes single user, no cooperation, lower bound. 'UB' denotes two user optimal joint processing upper bound.

and for our two user cooperation setting, the ergodic capacities are

$$\begin{aligned} C_{erg}(1) &= e^{1/P_s} E_1(1/P_s) \\ C_{erg}(2) &= 1 + e^{1/P_s} E_1(1/P_s) - 1/P_s e^{1/P_s} E_1(1/P_s). \end{aligned} \tag{10}$$

Figure 2 illustrates the lower and upper bounds of two cooperating users.

## V. Amplify Forward Cooperation

In what follows, we consider three types of cooperation schemes:

1) *Naive AF* - In this cooperation scheme the relaying user directly scales its input to the available transmit power $P_r$, and forwards the scaled channel output to the destination user using a single session $K = 1$. The destination then decodes the data based on its direct link channel output $y_1$ and the output of the cooperation link $y_{1,2}^{(1)}$.
2) *Separate preprocessing* - AF after removal of common layers that are separately decoded by each user. That is, each receiver attempts decoding on its own. Then both users exchange the index of the lowest layer decoded, and remove the commonly decoded signal from the channel output. Finally, the relaying user scales the residual signal to $P_r$, and forwards it to destination. This forms a single cooperation session ($K = 1$). The destination then optimally combines its own copy of the residual signal and the relayed version, and then decodes as many layers as possible.

3) *Multi-session* - multiple cooperation blocks ($K \to \infty$) with separate preprocessing per cooperation session, and a total power constraint $P_r$ for all the cooperation sessions. In order to maintain maximal average throughput, a wide-band cooperation link (4) is required. In this setting, common layers are removed before every AF session by both users, and after every AF transmission each user tries to decode more layers based on all received AF signals and its original received input signal. A closed form expression for the achievable rate is derived for unlimited number of sessions, assuming an overall power constraint $P_r$ for all sessions.

### A. Naive AF Cooperation

In the naive AF strategy, the relaying user ($i = 2$) scales its input to the available transmit power $P_r$, and forwards the signal to the destination user ($i = 1$). The received signal at the destination after AF is

$$\mathbf{y}_b = \begin{bmatrix} \mathbf{y}_{1,2}^{(1)} \\ \mathbf{y}_1 \end{bmatrix} = \begin{bmatrix} \alpha h_2 \mathbf{x}_s + \alpha \mathbf{n}_2 + \mathbf{w}_2 \\ h_1 \mathbf{x}_s + \mathbf{n}_1 \end{bmatrix} = \begin{bmatrix} \sqrt{\beta}\mathbf{x}_s + \widetilde{\mathbf{w}}_2 \\ h_1 \mathbf{x}_s + \mathbf{n}_1 \end{bmatrix} \tag{11}$$

where $y_b$ is the signal to be decoded at the destination, and the scaling factor $\alpha$ scales the transmit power to $P_r$, thus $\alpha = \sqrt{\frac{P_r}{1+P_s s_2}}$, where $s_i = |h_i|^2$. The normalized noise vector $\widetilde{\mathbf{w}}_2$ has i.i.d elements distributed $\mathcal{CN}(0,1)$, hence the normalized signal gain after the scaling of user $i = 2$ is

$$\beta = \frac{P_r s_2}{1 + P_s s_2 + P_r}. \tag{12}$$

The achievable rate as a function of the channel fading gains is given by the following mutual information

$$I(x_s; y_b|h_1, h_2) = \log(1 + P_s(s_1 + \beta)) = \log\left(1 + P_s\left(s_1 + \frac{P_r s_2}{1 + P_s s_2 + P_r}\right)\right). \tag{13}$$

Therefore the continuous broadcasting equivalent fading parameter is $s_b = s_1 + \beta$. This requires the derivation of the CDF of $s_b$, [25]

$$F_{s_b}(x) = Prob(s_b \le x) = \int_0^\infty du f_{s_1}(u) \int_0^{\max\left(0, x - \frac{P_r u}{1+P_s u+P_r}\right)} dv f_{s_2}(v), \tag{14}$$

where $f_{s_i}(u)$ is the PDF of $s_i$. For a Rayleigh fading channel with $f_{s_i}(u) = e^{-u}$ the CDF of $s_b$ is explicitly given by

$$F_{s_b}(x) = \begin{cases} 0 & x \leq 0 \\ 1 - e^{-\frac{(1+P_r)x}{P_r - P_s x}} - \int\limits_0^{\frac{(1+P_r)x}{P_r - P_s x}} du e^{-u-x+\frac{P_r u}{1+P_s u+P_r}} & 0 \leq x < \frac{P_r}{P_s} \\ 1 - \int\limits_0^{\infty} du e^{-u-x+\frac{P_r u}{1+P_s u+P_r}} & x \geq \frac{P_r}{P_s} \end{cases} \quad (15)$$

The corresponding PDF $f_{s_b}(x)$ is given by

$$f_{s_b}(x) = \begin{cases} 0 x \leq 0 \\ \int\limits_0^{\frac{(1+P_r)x}{P_r - P_s x}} du e^{-u-x+\frac{P_r u}{1+P_s u+P_r}} & 0 \leq x < \frac{P_r}{P_s} \\ \int\limits_0^{\infty} du e^{-u-x+\frac{P_r u}{1+P_s u+P_r}} & x \geq \frac{P_r}{P_s} \end{cases} \quad (16)$$

We can now state the outage and broadcasting achievable rates for the naive AF.

*1) Outage Approach:* Using the result of the fading power distribution (15)-(16), one can optimize for maximum average rate using a single level code. Since it has a SISO equivalent representation with fading distribution (15), the maximal average rate is

$$R_{out} = \max_{x>0} (1 - F_{s_b}(x)) \log(1 + x P_s), \quad (17)$$

where the transmitter uses code rate which is given by $\log(1 + x P_s)$. The rate $R_{out}$ can be evaluated numerically.

*2) Broadcast Approach:* In this approach the transmitter performs continuous code layering, matched to the equivalent fading random variable $s_b$ (from equation 14). Using the equivalent SISO channel model, and using the results of [25], the average received rate is given by

$$R_{AF,bs} = \max_{I(x), \text{ s.t. } I'(x) \leq 0} \int\limits_0^{\infty} (1 - F_{s_b}(x)) \frac{-x I'(x)}{1 + x I(x)} dx \quad (18)$$

where $f_{s_b}(x)$ is the PDF of $s_b$, the optimal residual interference distribution $I_{NAF}(x)$ is given by [25]

$$I_{NAF}(x) = \begin{cases} P_s & 0 \leq x \leq x_0 \\ I_r(x) & x_0 \leq x \leq x_1 \\ 0 & x \geq x_1 \end{cases} \quad (19)$$

where $I_r(x) \triangleq \frac{1-F_{s_b}(x)-xf_{s_b}(x)}{f_{s_b}(x)x^2}$, and $x_0$ and $x_1$ are determined from the boundary conditions $I_r(x_0) = P_s$ and $I_r(x_1) = 0$, respectively. Notice that $I_{NAF}(x)$ is indeed decreasing, starting from $P_s$ at $x = 0$. The average rate is explicitly given by

$$R_{NAF} = \int_0^\infty dx \left[\frac{2(1-F_{s_b}(x))}{x} + \frac{(1-F_{s_b}(x))f'_{s_b}(x)}{f_{s_b}(x)}\right]. \tag{20}$$

The first derivative of the PDF of $s_b$ is denoted by $f'_{s_b}(x)$.

### B. *Amplify Forward with Separate Preprocessing*

In this approach we assume that every user attempts decoding as many layers as possible independently, before the cooperation. Then both users exchange the index of the highest layer successfully decoded. Every user re-encodes the decoded data, up to the lower index (reconstructing only common information) and subtracts it from the original received signal. The relaying user scales the result into power $P_r$ and transmits over the cooperation link to the destination $i = 1$. This is better than the naive AF, since the cooperation is more efficient, resulting in higher equivalent gains. Like the naive AF, it requires only single session $K = 1$, but unlike the naive AF, it requires the knowledge of the destination fading in the relaying user. This strategy is directly matched with continuous broadcasting, as for every fading gain there is a different independent decoding capability. And for every decoding level there is an associated residual interference function. The received signal at the second user side can be expressed as follows,

$$\boldsymbol{y}_2 = h_2(\boldsymbol{x}_{s,D} + \boldsymbol{x}_{s,I}) + \boldsymbol{n}_2, \tag{21}$$

where $\boldsymbol{x}_{s,D}$ is the part of the source data successfully independently decoded by user $i = 2$. The residual interference signal is then denoted $\boldsymbol{x}_{s,I}$, which includes coded layers which were not decoded independently.

Assuming that $s_1 \geq s_2$, then the decoded data in $\boldsymbol{x}_{s,D}$ will include layers up to the parameter $s_2$. Let the residual interference power be denoted by $I(s)$, where $s$ is the fading gain equivalent. Thus after removing layers up to $s_2$ the residual interference power is given by $I(s_2)$. The residual signals at both sides (before a cooperation session) are then given by

$$\boldsymbol{y}_{1,I} = h_1\boldsymbol{x}_{s,I(s_2)} + \boldsymbol{n}_1. \tag{22}$$

$$\boldsymbol{y}_{2,I} = h_2\boldsymbol{x}_{s,I(s_2)} + \boldsymbol{n}_2. \tag{23}$$

It can be shown, following the same lines of AF derivation (12), that the equivalent fading gain, after amplifying and forwarding $y_{2,I}$, is

$$s_a = s_1 + \frac{P_r s_2}{1 + s_2 I(s_2) + P_r}. \tag{24}$$

In general, the cooperating user removes only common information from its input signal and forwards the residual signal to the destination. That is, each user tries decoding separately as many layers as possible. The destination user receives only new information when the helping user can decode at least the same number of layers. If the helping user had worse channel conditions it transmits its scaled residual interference, including layers which could be independently decoded by the destination. The equivalent fading gain observed by the destination, and its distribution are stated in the following proposition.

*Proposition 5.1:* In an *AF with separate preprocessing* cooperation strategy, with a single cooperation session $K = 1$ (power $P_r$), the highest decodable layer is associated with an equivalent fading gain determined by

$$s_a = s_1 + \frac{P_r s_2}{1 + s_2 \cdot \max(I(s_1), I(s_2)) + P_r}, \tag{25}$$

with the following CDF for a Rayleigh fading channel,

$$F_{s_a}(x) = \int_0^{\phi_1^{-1}(x)} \left(\exp\left(-2u\right) - \exp\left(-u - \phi_2(u)\right) - \exp\left(-u - \phi_3(u)\right)\right) du. \tag{26}$$

where

$$\begin{aligned} \phi_1(u) &= u + \tfrac{uP_r}{1+uI(u)+P_r} \\ \phi_2(u) &= \max\left(u, x - \tfrac{uP_r}{1+uI(u)+P_r}\right) \\ \phi_3(u) &= \max\left(u, \phi_4(x-u)\right) \end{aligned} \tag{27}$$

where

$$\phi_4(x-u) = \begin{cases} \frac{(1+P_r)(x-u)}{P_r - I(u)(x-u)} & P_r - I(u)(x-u) > 0 \\ \infty & P_r - I(u)(x-u) <= 0 \end{cases} \tag{28}$$

*Proof:* See Appendix A.

Note that in the AF with separate preprocessing strategy, we have implicitly assumed that if both users can decode all layers independently, then no forwarding is done. This saves a fraction of the relaying power $P_r$, and under long-term power constraint on the relay, AF transmission power may be increased by $\frac{1}{1-P_{bs}}$, where $P_{bs}$ is the probability that both users will successfully decode all layers.

The expressions for the broadcasting average rate include the function $I(s)$ as part of the equivalent fading gain CDF $F_{s_a}(x)$, and in an integral form. This turns the optimization problem of the average rate to be a difficult one. And it seems that no closed analytical solution for optimal $I(s)$ can be found. We suggest few sub-optimal approaches to maximize the achievable average broadcasting rate:

1) *One step sub-optimal* $I_{sub-opt}(s)$. Use a sub-optimal power allocation, $I_{NAF}(s)$, which is the optimal naive AF power allocation (19) to compute the corresponding CDF of the equivalent fading gain (26). Then use these distributions to compute $R_{bs,sub-opt}$.
2) *Iterative solution of* $I(s)$. Assume at the first iteration that $I_0(s)$ is given by the naive AF function specified in (19). Calculate $F_{s_a,1}(x)$ using $I_0(s)$, and compute the corresponding average rate $R_{1,bs}$. In the second iteration, calculate $I_1(s)$ using $F_{s_a,1}(x)$ and (19). Go back to (25) and solve for $F_{s_a,2}(x)$ using $I_1(s)$, and compute $R_{2,bs}$. Repeat the same procedure till the difference $|R_{k,bs} - R_{k-1,bs}|$ is sufficiently small.
3) *Finite level coding*. The derivation of $F_{s_a}(x)$ in this case is doable, and the maximal average rates may be numerically computed. Although with two level coding, the efficiency of separate preprocessing may be very limited, since there are only two thresholds involved. So that separate preprocessing may help only when both users successfully decode the first layer, and could not decode the second layer.

### C. Multi-Session Amplify and Forward with Separate Preprocessing

We consider here the multi-session AF with separate preprocessing per session. The total power allocation available for all sessions is $P_r$, where unlike previous schemes, here $K = \infty$. In this setting, common layers are subtracted before every AF session by both users, and after every AF transmission each user tries to decode more layers based on all received AF signals and its original received input signal. We find the average rate for unlimited number of sessions, assuming only an overall power constraint for all sessions. It should be emphasized that the multi-session is performed over parallel channels (for example, OFDM), as illustrated in Figure 3, in such way that the source transmission is block-wise continuous. For example, during the $k^{th}$ cooperation session of the $1^{st}$ transmitted block (from the source), the $1^{st}$ cooperation session for the $k-1$ transmitted block takes place. As the overall multi-session power is limited to $P_r$, at every block epoch a total power of $P_r$ is used.

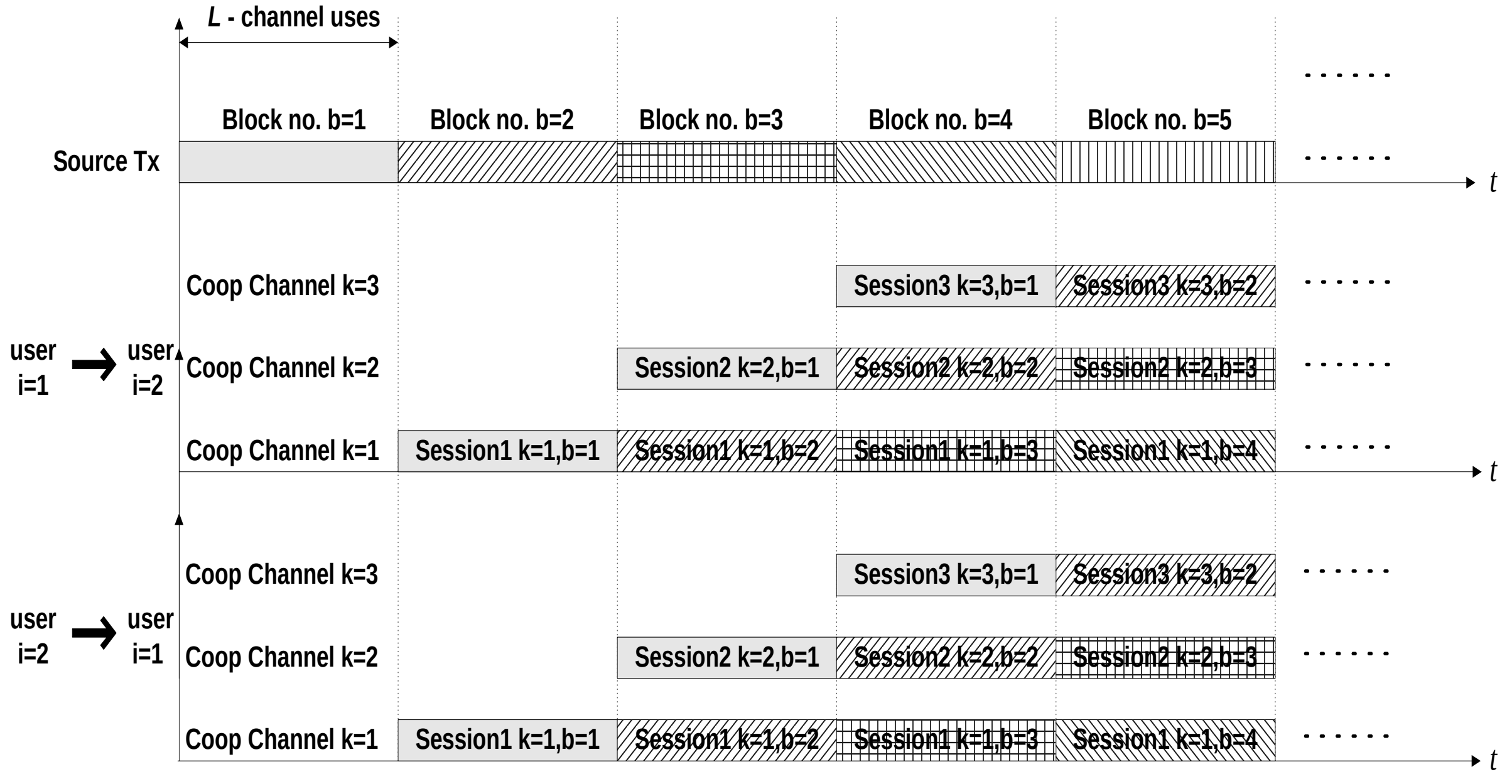

Fig. 3. Illustration of multi-session AF cooperation with $K = 3$ cooperation sessions per block. Block $b$ refers to the $b^{th}$ transmission for which there is a fixed fading level. The source transmits continuously information blocks, and simultaneous cooperation sessions take place on parallel channels.

Since the cooperation is performed over parallel channels, with infinitesimal power $\rho(s)$ allocated per channel, the average capacity of this wide-band cooperation link is the capacity of the corresponding parallel channel. The power allocation constraint enforces $\int_0^\infty \rho(s)ds = P_r$. The fractional rate per sub-band is then $dR(s) = \log(1+\rho(s)ds) = \rho(s)ds$, [33]. Therefore, the average capacity of this wide-band cooperation link, regardless of the actual power allocation density is $C_{coop} = P_r$ (4). Notice that we use AF, which can not effectively use such capacity increase in a single session cooperation ($P_r > \log(1+P_r)$).

In the case of unlimited sessions, the scalar equivalent fading gain can be derived for a given broadcasting power allocation $I(s)$. From the equivalent fading gain a CDF can be computed, from which the average achievable rate can be obtained.

*Proposition 5.2:* In a *multi-session AF ($K \to \infty$, cooperation power constraint $P_r$) with separate preprocessing* cooperation strategy, the highest decodable layer is associated with an equivalent fading gain determined by

$$s_{ms} = \begin{cases} s_a^* & s_1 \geq s_2 \\ s_b^* & s_1 < s_2 \end{cases} \quad (29)$$

where $s_b^*$ is the solution of the following equation,

$$\int_{s_2}^{s_b^*} \frac{s_1}{(s_1+s_2-\sigma)^2}[1+s_1 I(\sigma)]d\sigma = P_r, \quad (30)$$

and by using $s_b^*$,

$$s_a^* = s_1 + s_2 \frac{Z(s_b^*)}{1+Z(s_b^*)}. \quad (31)$$

where

$$Z(s) = \int_{s_2}^{s} \frac{1+s_1 I(\sigma)}{(1+s_2 I(\sigma))} \frac{s_1}{(s_1+s_2-\sigma)} d\sigma \quad (32)$$

*Proof:* See Appendix B.

For a given power allocation $I(s)$ computation of the CDF of $s_{ms}$ is quite involved, as it requires solving (30) for every pair $(s_1, s_2)$ subject to $s_1 \geq s_2$. Hence the optimization of $I(s)$ to maximize the achievable rate does not seem doable. For the numerical results we use $I(s)$ corresponding to optimal broadcasting in presence of optimal joint decoding. This selection is demonstrated (see Section VIII) to be a good one, particularly for high $P_s$ and $P_r$, as such conditions allow approximation of optimal performance with multi-session AF cooperation.

We have used a continuous power allocation function for the multi-session cooperation link power $\delta(s)$ (see appendix B), such that for every session, different power may be used ($s$ here serves as a continuous session index). Identical $\delta(s)$ for both cooperation directions are used to simplify derivation, although such restriction is suboptimal, since $\delta(s)$ is chosen to maximize $s_b^*$, which is not equal to averaged rate (which includes also $s_a^*$). In addition, the scheme is suboptimal by letting user $i = 1$ forward layers from $s_b^{(k)}$, rather than $s_a^{(k)}$, so it forwards a layered transmission instead of a direct transmission (which is more efficient, since the cooperative channel is non-fading).

Notice that both $s_a^*$ and $s_b^*$ reach $s_1 + s_2$ when $P_r \to \infty$, which is the same case with the other $AF$ approaches. The difference, however, is the convergence rate to $s_1 + s_2$ of the different cooperation schemes. This is demonstrated in the numerical results section VIII.

## VI. Compress Forward Cooperation

In this section we consider compress forward cooperation. Both users are capable of quantizing and compressing their received signals and forwarding the result to one another. The compression here relies on the well known Wyner-Ziv [29] compression using side information at the decoder. Similar to the AF, here too, we consider three ways of implementing the basic cooperation.

### A. Naive CF Cooperation

Consider the channel model in (1)-(2). The signal to be sent to the second user $\hat{\boldsymbol{y}}_1$, is compressed in the Wyner-Ziv spirit, and is given by

$$\hat{\boldsymbol{y}}_1 = \boldsymbol{y}_1 + \boldsymbol{n}_c = h_1 \boldsymbol{x}_s + \boldsymbol{n}_1 + \boldsymbol{n}_c, \tag{33}$$

where $n_c \sim \mathcal{CN}(0, \sigma^2)$ is the compression noise, which is independent of $y_1$. Then the maximal achievable rate, for the second user, is given by

$$\begin{aligned} &R_{WZ,2}(h_1, h_2) = I(x_s; y_2, \hat{y}_1 | h_1, h_2) \\ &\text{s.t.} \quad I(y_1; \hat{y}_1 | h_1) - I(y_2; \hat{y}_1 | h_1, h_2) \leq C_{coop} \end{aligned} \tag{34}$$

where $R_{WZ,2}(h_1, h_2)$ is maximized when the constraint is met with equality. The constraint $C_{coop}$ represents the cooperation link capacity. According to the channel model there is unlimited bandwidth and a $P_r$ power limitation. We consider two cases for the naive CF:

1) *Narrow-band naive CF* - In this case the cooperation bandwidth is equal to the source-relay link bandwidth ($K = 1$), and therefore the cooperation capacity is $C_{coop} = \log(1 + P_r)$.
2) *Wide-band naive CF* - In this case the cooperation bandwidth is unlimited ($K = \infty$), and according to (4) the cooperation capacity is $C_{coop} = P_r$, when fractional power is allocated per sub-band.

When requiring that the constraint in (34) will be met by equality, with a narrow-band cooperation link, the resulting $\mathrm{E}|n_c|^2 = \sigma^2$ is (C.7),

$$\sigma^2_{NB} = \frac{1 + s_1 P_s + s_2 P_s}{P_r (1 + s_2 P_s)} \tag{35}$$

*Proposition 6.1:* In a *Narrow-band Naive Wyner-Ziv compression* cooperation strategy, the highest decodable layer is associated with an equivalent fading gain determined by

$$s_{NWZ} = s_2 + \frac{s_1 (1 + s_2 P_s) P_r}{(1 + P_r)(1 + s_2 P_s) + s_1 P_s}. \tag{36}$$

The distribution of $s_{NWZ}$ for a Rayleigh fading $F_{s_{NWZ}}(u)$ is, when $u \geq \frac{P_r}{P_s}$:

$$F_{s_{NWZ}}(u) = 1 - e^{-u}\left(1 + \frac{P_r(uP_s+1)}{P_s(P_r+1)}\right) + \frac{P_r(uP_s+1)^2}{P_s^2(P_r+1)^2}e^{-\frac{uP_rP_s-1}{P_s(P_r+1)}}\mathrm{Ei}\left(1, \frac{uP_s+1}{P_s(P_r+1)}\right) \quad (37)$$

and when $u < \frac{P_r}{P_s}$:

$$F_{s_{NWZ}}(u) = 1 - e^{-u}\left(1 + \frac{P_r(uP_s+1)}{P_s(P_r+1)}\right) - \frac{P_r - uP_s}{(1+P_r)P_s}e^{-\frac{u}{P_r-uP_s}} + \frac{P_r(uP_s+1)^2}{P_s^2(P_r+1)^2}e^{-\frac{uP_rP_s-1}{P_s(P_r+1)}}$$
$$\left(\mathrm{Ei}\left(1, \frac{uP_s+1}{P_s(P_r+1)}\right) - \mathrm{Ei}\left(1, \frac{P_r(uP_s+1)^2}{P_s(P_r+1)(P_r-uP_s)}\right)\right). \quad (38)$$

In the same lines of derivation for the narrow-band cooperation, when requiring that the constraint in (34) will be met by equality, for a wide-band cooperation link ($C_{coop} = P_r$), the resulting $\mathrm{E}|n_c|^2 = \sigma^2$ is (C.8),

$$\sigma_{WB}^2 = \frac{1 + s_1P_s + s_2P_s}{(e^{P_r}-1)(1+s_2P_s)} \quad (39)$$

where it may be noticed that in a wide-band cooperation regime the noise variance of the compressed signal decays exponentially fast with $P_r$.

### B. Wyner-Ziv compress and forward with separate preprocessing

Let us repeat what was done for the Amplify and Forward with separate preprocessing in subsection V-B, for Wyner-Ziv compression. For consistency, assume that $s_1 > s_2$, and then replace $P_s$ by $I(s_2)$ in (35), by introducing the preprocessing, and letting the receivers subtract the decoded message before compressing and forwarding. We get that (35) is now

$$\sigma^2 = \frac{1 + s_2I(s_2) + s_1I(s_2)}{P_r(1 + s_1I(s_2))} \quad (40)$$

and the equivalent signal to noise ratio at $i = 1$, after the first iteration is now written by (36) and (40),

$$s_a^{(1)} = s_1 + \frac{s_2P_r(1 + s_1I(s_2))}{(1+P_r)(1+s_1I(s_2)) + s_2I(s_2)}. \quad (41)$$

### C. Multiple sessions with Wyner-Ziv compression and separate preprocessing

As was done for the amplify and forward, can be repeated for the Wyner-Ziv processing. For this to be performed, several definitions are in order. Notice that each step of Wyner-Ziv compression can use all information collected in the previous sessions, in the form of side information.

Define $\hat{\boldsymbol{y}}_1^{(k)} = \boldsymbol{y}_1 + \boldsymbol{n}_{c,1}^{(k)}$, where $n_{c,1}^{(k)}$ is independent of $y_1$, as the compressed signal that is transmitted from $i = 1$ to the colocated user, $i = 2$. We refer the reader to [34], for successive Wyner-Ziv overview. Here, we deal with the case where the message that is transmitted in each session has better side information than the previous session, since more layers are decoded. Further, the second session can use the information sent by all the previous sessions, in order to improve performance. Since the power that is used by each session is a control parameter, rather than a fixed parameter, the use of an auxiliary variable that is transmitted during a session, but decoded only at the next session (due to the better side information, declared as $V$ in [34]) is superfluous. Next, using [34], the following Markov chain is defined, where unlike [34], we are interested in independent averaged distortion, rather than plain averaged distortion.

$$y_2 - x_s - y_1 - \hat{y}_1^{(k)} - \hat{y}_1^{(k-1)} - \cdots - \hat{y}_1^{(1)} \tag{42}$$

$$y_1 - x_s - y_2 - \hat{y}_2^{(k)} - \hat{y}_2^{(k-1)} - \cdots - \hat{y}_2^{(1)} \tag{43}$$

The equivalent fading gains after every iteration of the multi-session cooperation are stated in the following proposition.

*Proposition 6.2:* The achievable rate in the multi-session with separate preprocessing and successive refinement WZ is give in a recursive form for the $k^{th}$ session,

$$R_{WZ}^{(k)} = E_{s_{ms}^{(k)}} \log(1 + s_{ms}^{(k)} P_s) \tag{44}$$

where

$$s_{ms}^{(k)} = \begin{cases} s_a^{(k)} & s_1 \geq s_2 \\ s_b^{(k)} & s_1 < s_2 \end{cases} \tag{45}$$

and

$$s_a^{(k)} = s_1 + \frac{s_2}{1 + (\sigma_2^{(k)})^2} \tag{46}$$

$$s_b^{(k)} = s_2 + \frac{s_1}{1 + (\sigma_1^{(k)})^2}, \tag{47}$$

and

$$\left(\sigma_j^{(k)}\right)^2 = \left(\sigma_j^{(k-1)}\right)^2 \frac{1 + s_j I(s^{(k-1)}) + s_{3-j} I(s^{(k-1)})}{(1 + s_{3-j} I(s^{(k-1)})) \left[1 + \delta_j^{(k)} \left(1 + \left(\sigma_j^{(k-1)}\right)^2\right)\right] + s_j I(s^{(k-1)})(1 + \delta_j^{(k)})} \tag{48}$$

for $j = 1, 2$, and where $\delta_j^{(k)}$ is the fractional power assigned to user $j$ for the $k^{th}$ cooperation session.

*Proof:* See Appendix D.

## VII. Decode Forward Cooperation

We consider here the well known form of cooperation, namely Decode and Forward (DF). We present here bounds for the DF strategy, where the clear upper bound is the strongest user achievable rate (similar to selection diversity). From the nature of this approach, there is no place for considering multi-session, as after one session there is nothing the destination can send back to the relay for improving upon its independent decoding. For a fair comparison of DF cooperation to other multi-session techniques we consider both wide-band cooperation, where $C_{coop} = P_r$ (4), and narrow-band cooperation (corresponding to the single session relaying techniques), where the cooperation link capacity is only $C_{coop} = \log(1 + P_r)$ (3).

The DF strategy may be described as follows. The source performs continuous broadcasting, and two copies of the transmitted signal are received at destination and relaying side, as described by the channel model (1)-(2). Recalling that the destination is denoted by user $i = 1$, then for $s_1 \geq s_2$ the destination user can decode at least as many layers as the relaying user. Hence there is place for DF cooperation **only** when $s_1 < s_2$, as in this case the relaying user can decode more layers than the destination. The **additional** layers decoded by the relay (for $s \in (s_1, s_2]$) are encoded by the relay and forwarded, constrained by the cooperation channel capacity. Thus for $P_r >> P_s$, a practically unlimited cooperation channel all additional information may be sent to destination and the strongest user upper bound is obtained.

Notice that for a fair comparison of DF to multi-session AF, wide-band cooperation is also considered. For wide-band cooperation the cooperation link capacity is $C_{coop} = P_r$ (4). While in the narrow-band cooperation, this link capacity is $C_{coop} = \log(1 + P_r)$.

Denote the decodable rate associated with a fading gain $s$ by $R(s)$, where $R(s) = \int_0^s du \frac{\rho(u)u}{1+I(u)u}$. Say that before cooperation starts user $i$ decodes $R(s_i)$. As mentioned cooperation is required for $s_1 < s_2$, and is limited by the relay link capacity $C_{coop}$. Hence for the pair $(s_1, s_2)$, the achievable broadcasting rate is given by

$$R_{DF}(s_1, s_2) = \begin{cases} \min\{R(s_1) + C_{coop},\ R(s_2)\} & s_2 > s_1 \\ R(s_1) & \text{otherwise} \end{cases} . \tag{49}$$

The optimal broadcasting power distribution maximizes the average rate, and the optimization problem is stated as follows,

$$\begin{aligned} \overline{R}_{DF} &= \max_{\rho(s)\geq 0,\ \text{s.t.} \int_0^\infty ds\rho(s)\leq P_s} E_{s_1,s_2} R_{DF}(s_1, s_2) \\ &= \max_{\rho(s)\geq 0,\ \text{s.t.} \int_0^\infty ds\rho(s)\leq P_s} \int_0^\infty ds_2 \int_0^{s_2} ds_1 f(s_1) f(s_2) \min\{R(s_1) + C_{coop},\ R(s_2)\} \\ &\quad + \int_0^\infty ds_2 \int_{s_2}^\infty ds_1 f(s_1) f(s_2) R(s_1) \end{aligned} \tag{50}$$

where $\rho(s) = -\frac{d}{ds}I(s)$ is the power density function. Finding the optimal power allocation seems intractable analytically, however $\overline{R}_{DF}$ could be computed for sub-optimal power distributions, such as the strongest user optimal $I_{sel,opt}(s)$, or for the no cooperation lower bound $I_{SU,opt}(s)$, and for $I_{Joint,opt}(s)$. These are defined and demonstrated in section VIII.

## VIII. Numerical Results

In this section, we compare the broadcasting and outage achievable rates of the various cooperation methods, with narrow band cooperation links for all schemes, besides the multi-session. Figures 4-5 demonstrate by numerical results the broadcasting AF and CF cooperation gains. Average achievable rates are computed for AF cooperation with a single session, which we have referred to as naive AF cooperation, and separate preprocessing. For the separate preprocessing we have used a sub-optimal power allocation which admits the optimal power allocation of naive AF broadcasting. Thus, in both cases we have the same power allocation, only in the latter, common information is removed prior to relaying. It may be noticed that when the SNR on the cooperation link satisfies $P_r \geq P_s$, the achievable rates are close to the joint processing upper bounds, where separate preprocessing is slightly better compared to the naive AF. However, when $P_r < P_s$ the separate preprocessing can introduce substantial gains over the naive AF. Note also that the computed separate preprocessing rate is yet a lower bound, since the optimal power allocation was not obtained. Separate preprocessing AF surpasses the outage upper bound (joint processing with an outage approach) for high SNRs. For example, in Figure

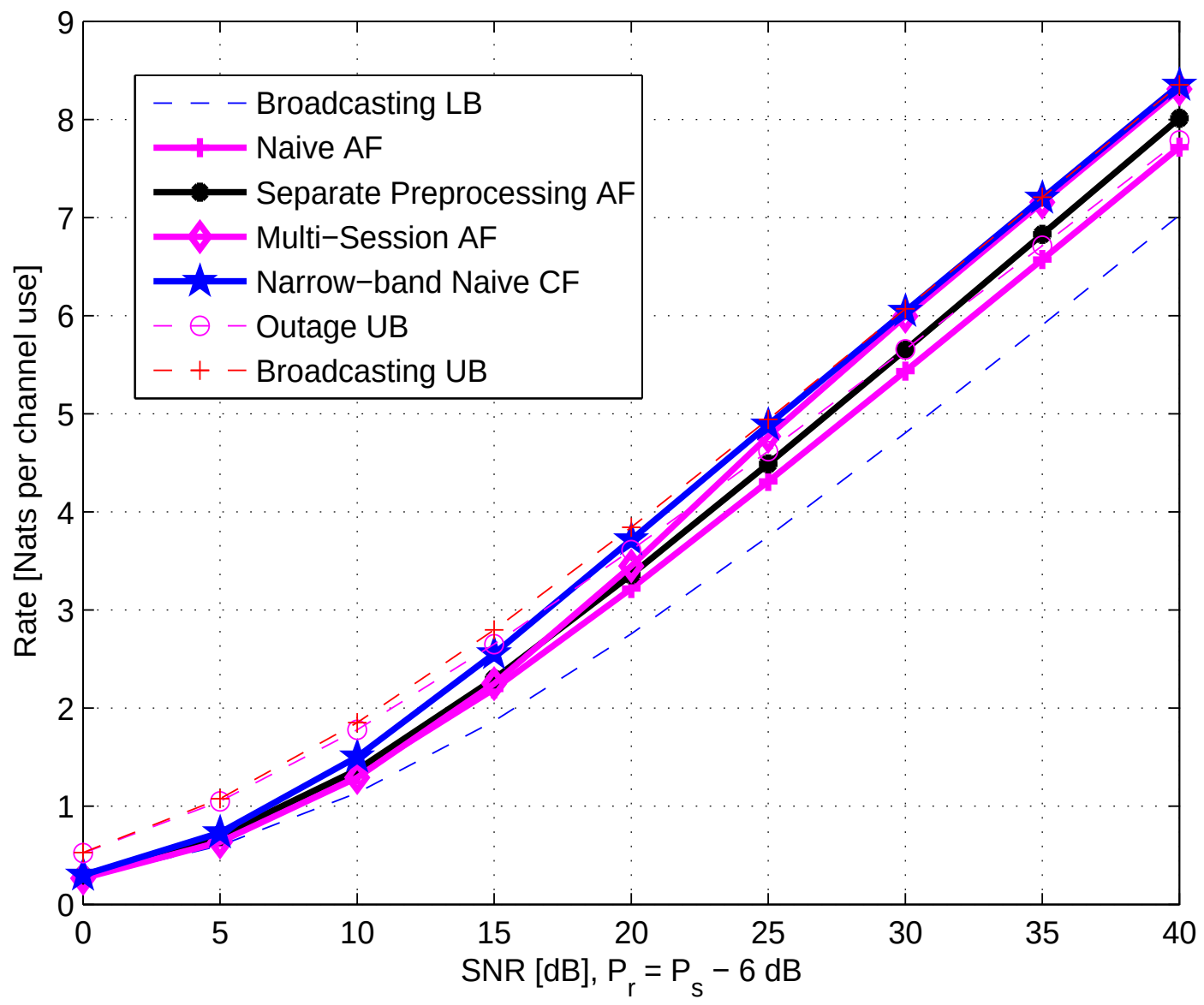

Fig. 4. Broadcast approach: average rates of Naive AF, AF with separate preprocessing, multi sessions AF and narrow-band (NB) naive CF compared to upper and lower bounds ($P_r = P_s - 6$ dB).

4, where $P_r = P_s - 6$ dB, the separate preprocessing AF achieves a $\sim 1$ dB gain over the outage upper bound. The multi-session achievable rates are computed using proposition 5.2, for the broadband cooperation channel ($K \to \infty$). The sub-optimal power distribution function $I(s)$ used for the rate computation is the one corresponding to the broadcasting upper bound (7), which is $I_{Joint,opt}(s) = \frac{1}{s^3} + \frac{1}{s^2} - \frac{1}{s}$. Interestingly, the average achievable rates with multi-session, with a sub–optimal power allocation approximate the broadcasting upper bound, for moderate and high SNRs, and for both $P_r/P_s = -6, 0$ dB ratios in Figures 4-5. Another efficient approach is the narrow-band naive CF which uses the Wyner-Ziv (WZ) compression based cooperation. This approach seems to be the best approach out of all considered settings. The naive WZ cooperation even closely approximates the separate processing WZ cooperation, as will be demonstrated in the following.

Figures 6-7 show a comparison between the naive AF, separate preprocessing AF, multi-session AF, and narrow-band naive CF, as function of the cooperation link quality ($P_r/P_s$). As may be noticed from these figures, the lower $P_r/P_s$, the higher the rate gains of separate preprocessing AF, over the naive approach. For $P_s = 20$ dB, both approaches achieve gains over

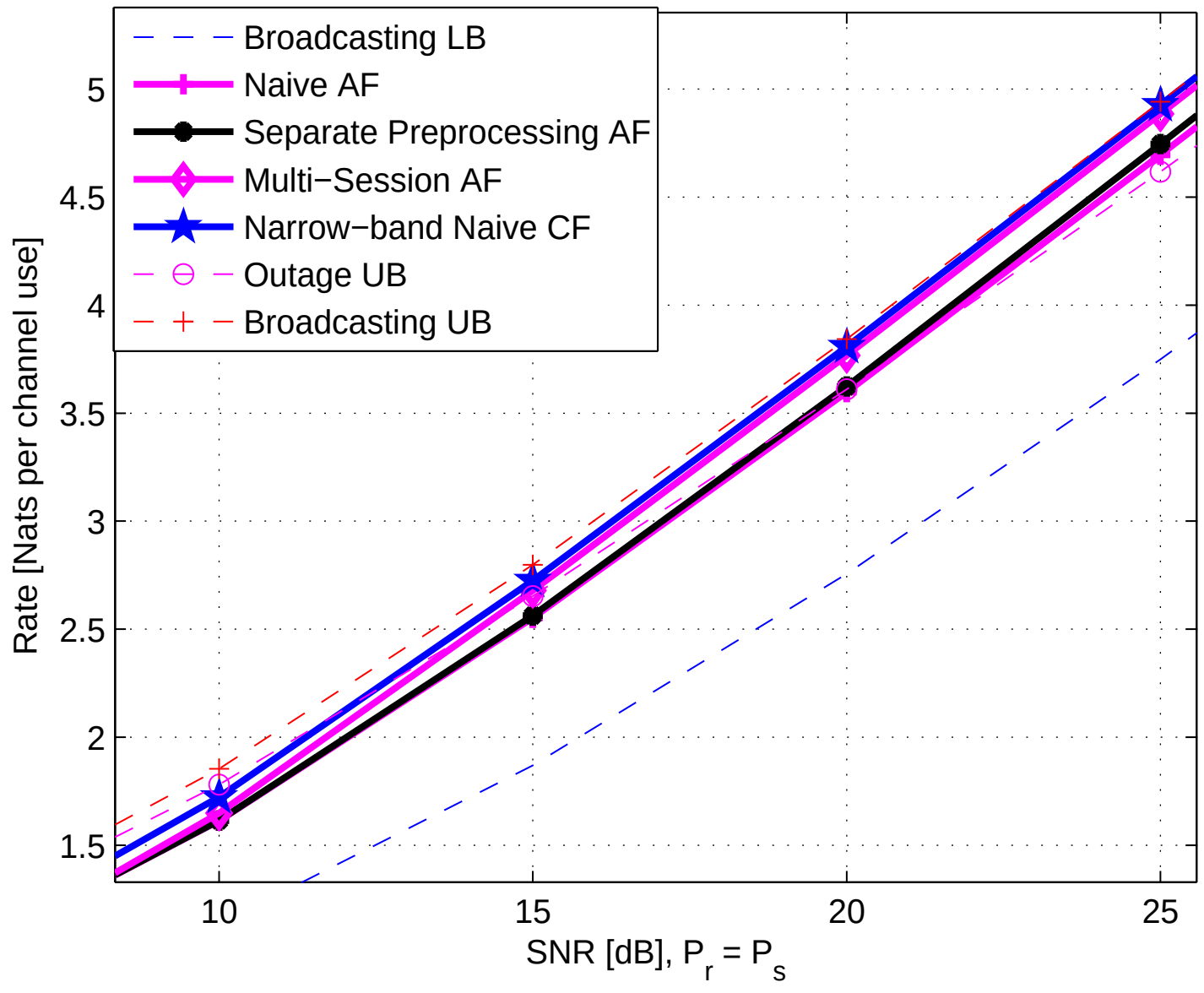

Fig. 5. Broadcast approach: average rates of Naive AF, AF with separate preprocessing, multi sessions AF and narrow-band (NB) naive CF compared to upper and lower bounds ($P_r = P_s$).

the outage upper bound for $P_r/P_s \geq 0$ dB. However, for $P_s = 40$, the separate preprocessing AF has increasing gains over the outage upper bound for any $P_r/P_s \geq -12$ dB. In view of the multi-session AF in Figures 6-7, it seems that for moderate to high $P_s$ and $P_r$, the multi-session AF approximates the broadcasting upper bound. The naive CF, again, outperforms all other approaches, and approximates the broadcasting upper bound even on a wider range of $P_r$ values.

Figure 8 demonstrates the implications of using sub-optimal power allocation for broadcasting in the AF multi-session and the narrow-band naive CF approaches. It may be noticed that for $P_r/P_s > -5$ dB it is more efficient to use the full cooperation optimal $I(s)$, however for lower relaying power values it is already preferable to use the single user optimal broadcasting power allocation. In the narrow-band naive CF approach the full cooperation optimal power distribution $I_{Joint,opt}(s)$ is highly efficient and approximates well the throughput with an optimal power allocation derived from the WZ approach. However, in the low $P_r/P_s$ values both power allocations show close performance to that with the single user optimal power allocation.

Figures 9-10 demonstrate achievable rates of the DF approach with single session only (as there

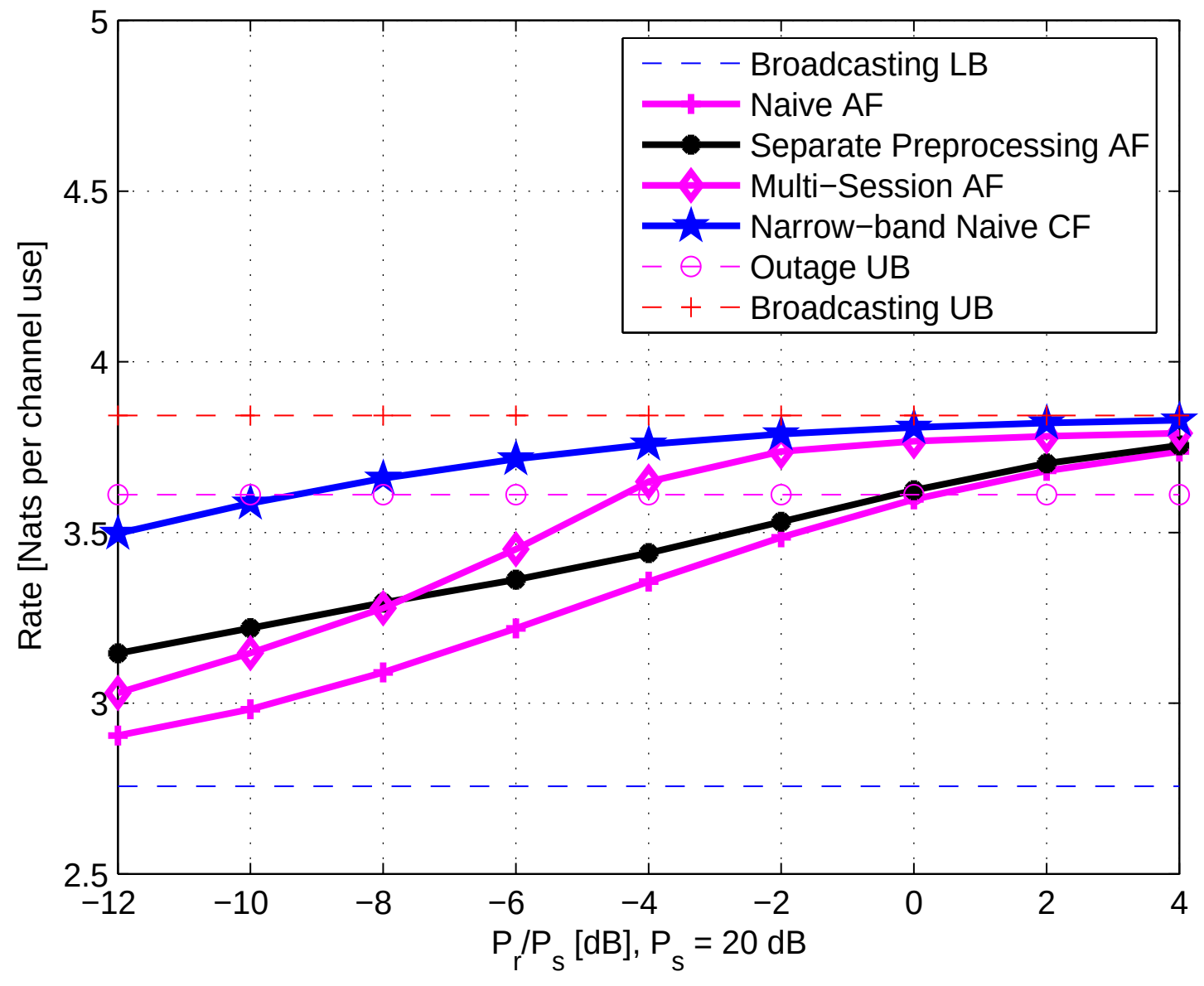

Fig. 6. Broadcast approach: average rates of Naive AF, AF with separate preprocessing, multi sessions AF and narrow-band (NB) naive CF compared to upper and lower bounds, as function of the channels quality ratio $\frac{P_r}{P_s}$. ($P_s = 20$ dB).

is no place for multi-session cooperation with a DF strategy only). Achievable rates are computed using (50) for narrow-band and wide-band cooperation channel link. The difference between the achievable rates is the power allocation strategy, which is sub-optimal for all three. This is since the optimal power distribution $I(x)$ for (50) is a difficult problem to solve analytically. The power allocations considered include $I_{SU,opt}(x)$ - single user optimal distribution; $I_{Joint,opt}(x)$ - full-cooperation joint decoding optimal distribution; and $I_{sel,opt}(x)$ is the optimal power distribution for strongest user (DF) upper bound. As may be noticed from the figures the best achievable rate of narrow-band cooperation uses $I_{sel,opt}(x)$, which also closely approximates the DF upper bound for $P_r \geq P_s$. For low SNRs and $P_r << P_s$ the achievable rate with $I_{SU,opt}(x)$ is slowest decaying, and will naturally be preferable in the extreme case of low $P_r$ (which in the limit is the case of no effective cooperation). Additionally, over a wide-band cooperation link the DF cooperation closely approximates the DF upper bound in all considered $P_r/P_s$ ratios and SNRs.

Figures 11-12, demonstrate achievable rates of the separate processing WZ cooperation scheme. As the optimal broadcasting power distribution does not lend itself to an analytical solution,

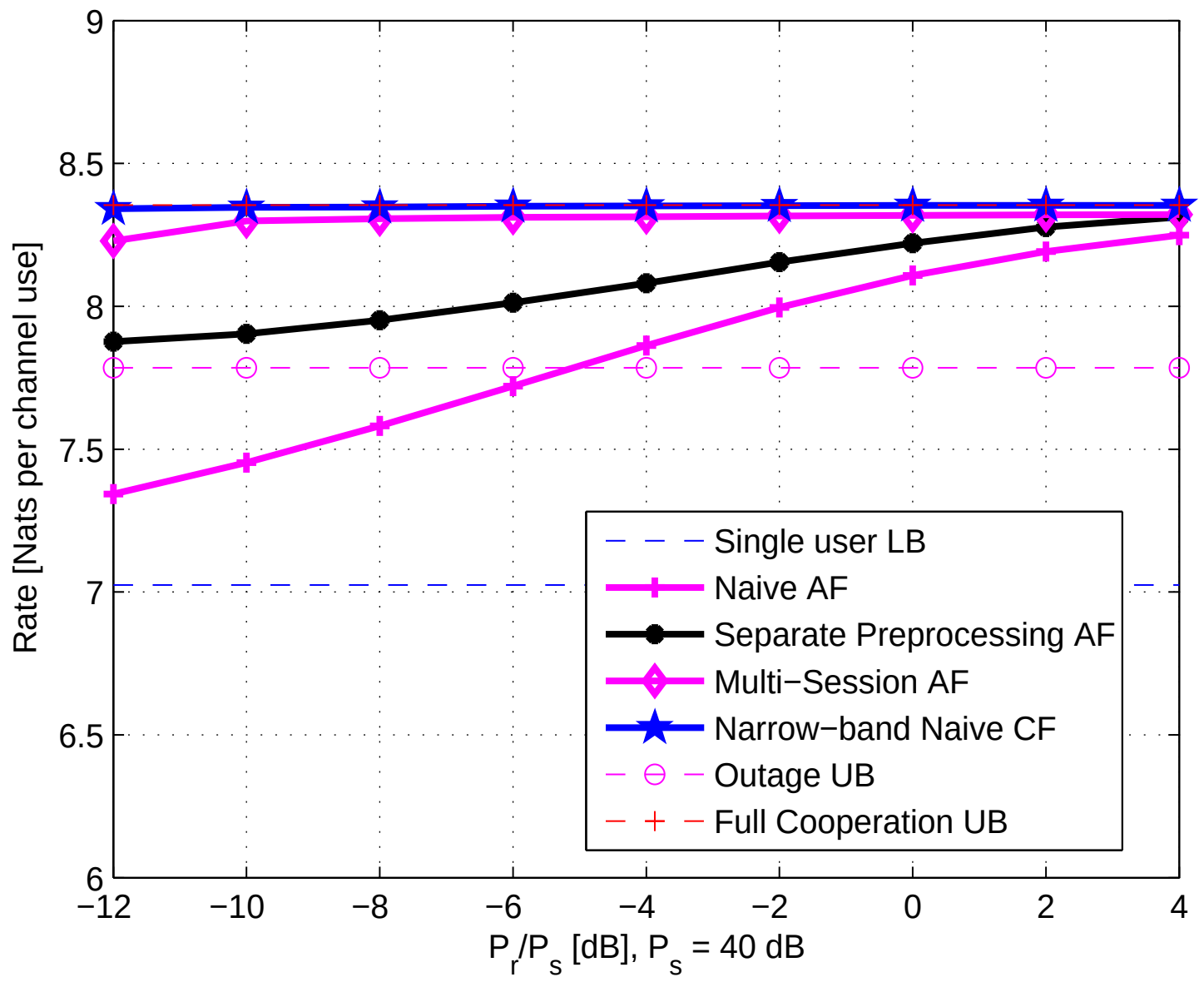

Fig. 7. Broadcast approach: average rates of Naive AF, AF with separate preprocessing, multi sessions AF and narrow-band (NB) naive CF compared to upper and lower bounds, as function of the channels quality ratio $\frac{P_r}{P_s}$. ($P_s = 40$ dB).

sub-optimal power distributions are used. The power distributions used are $I_{Joint,opt}(x)$ and $I_{NWZ,opt}(x)$. The function $I_{Joint,opt}(x)$ is the optimal power allocation for a broadcasting with optimal joint decoding. This function is expected to closely approximate the optimal power allocation of separate processing WZ for $P_r \geq P_s$, and high SNRs, which is also the case where the naive WZ cooperation closely approximates the broadcasting upper bound (see also Figure 8). The function $I_{NWZ,opt}(x)$ is the optimal power distribution for naive WZ cooperation. As may be noticed from Figures 11-12, the separate processing WZ with these sub-optimal power distributions gains only marginally compared to the naive WZ cooperation. The largest evident gain is for $P_s = 20$ dB, and for $P_r << P_s$ (Figure 11) with $I_{NWZ,opt}(x)$. Using $I_{Joint,opt}(x)$ in these cases may turn out to be even less efficient than naive WZ cooperation. For $P_s = 40$ dB it may be noticed that the separate processing WZ gain is negligible. These results indicate that naive WZ is already highly efficient, and that separate processing might provide significant gains, however we are unable to fully justify this, since the optimal power distribution for separate processing WZ is unknown.

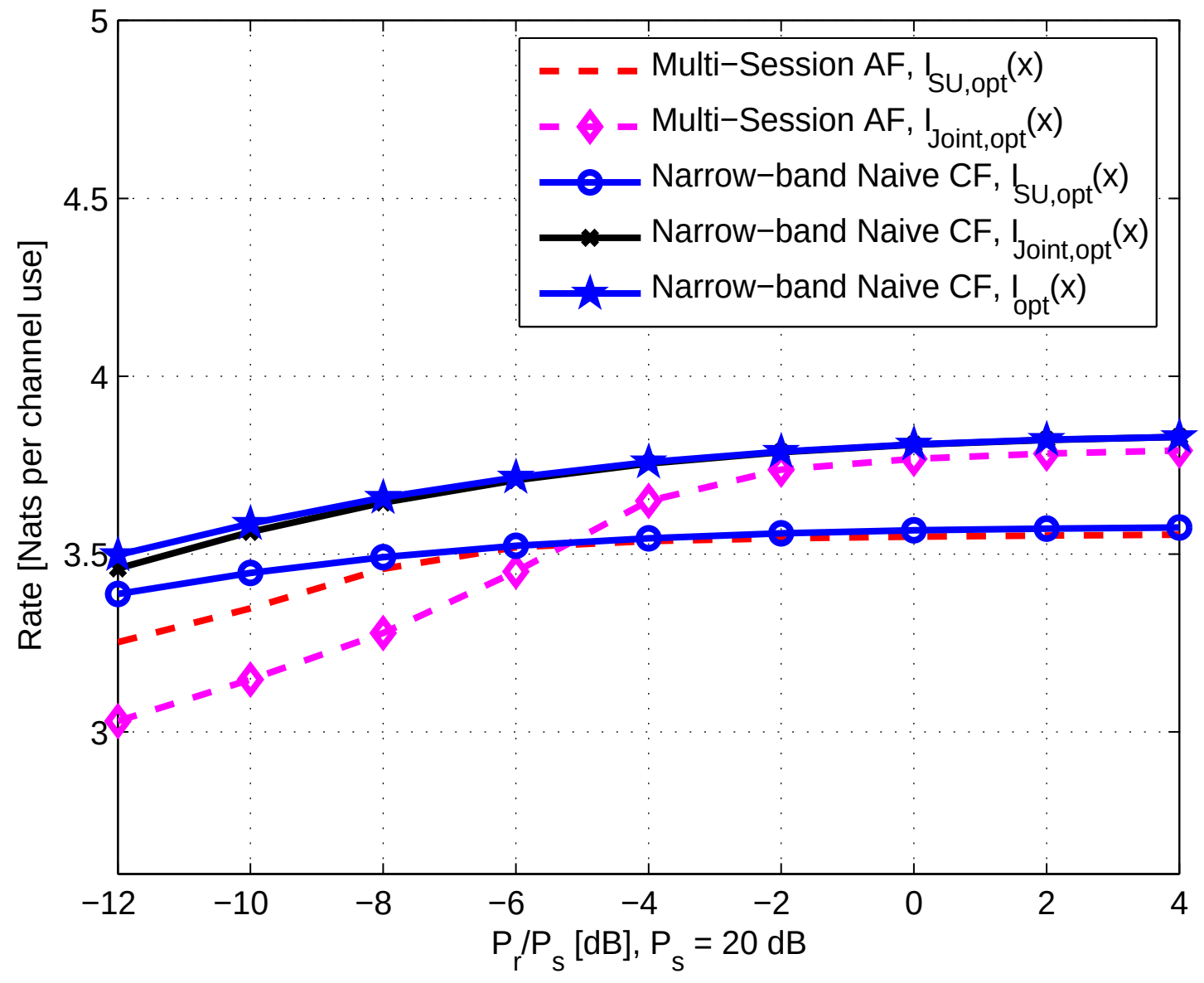

Fig. 8. Broadcast approach: average rates of multi sessions AF and narrow-band (NB) naive CF, as function of the channels quality ratio $\frac{P_r}{P_s}$. The approaches are compared as function of the broadcasting power allocation function, where $I_{SU,opt}(x)$ refers to the single user optimal power allocation, and $I_{Joint,opt}(x)$ denotes the function corresponding to full-cooperation joint decoding bound, and $I_{opt}(x)$ is the naive WZ optimal power allocation ($P_s = 20$ dB).

## IX. Discussion

We have considered two relaying techniques - AF and CF, for a multi-session cooperation, when the transmitter employs a broadcast transmission approach. Our cooperation strategies are designed to enhance the overall throughput of a destination user, while a colocated user receives another copy of the original transmitted signal over an independent fading channel. Essentially, the results here are also valid for the case when a single source sends common information for two users, and they cooperate following the described schemes as to maximize their individual throughput. One may consider the following communication schemes, among others:

1) **Information enhancement** - in this case two users are receiving common information such as digital TV broadcasting, and the cooperation allows for image quality enhancement. Progressive transmission of images is also a useful application here [35],[36], where refinement of image quality is achieved through decoding of more coded layers. In this transmission scheme, no acknowledge (ACK) signal to the transmitter is required.

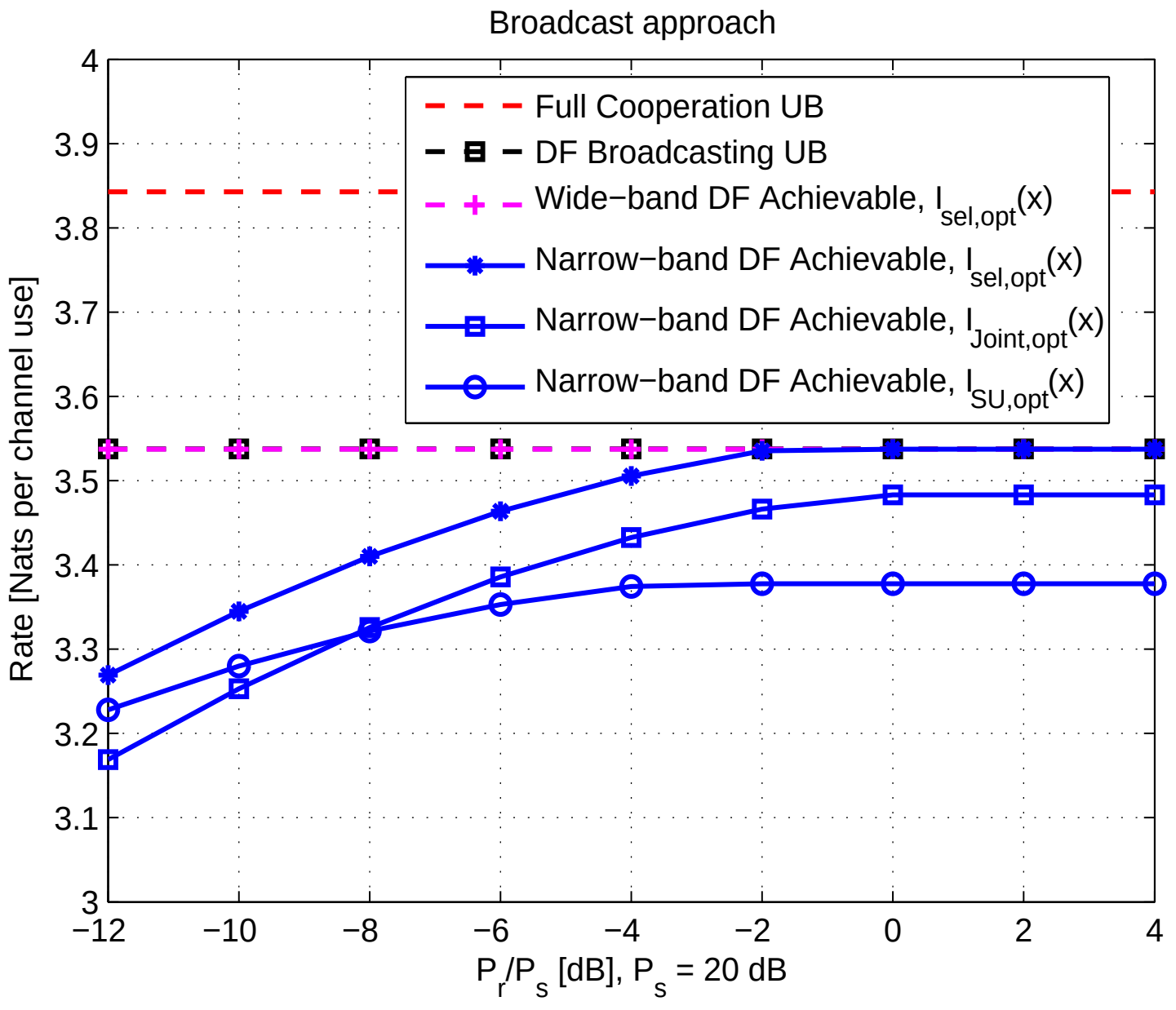

Fig. 9. Broadcast approach: average rates of DF achievable approaches, as function of the channels quality ratio $\frac{P_r}{P_s}$. The approaches are compared as function of the broadcasting power allocation function, where $I_{SU,opt}(x)$ refers to the single user optimal power allocation, and $I_{Joint,opt}(x)$ denotes the function corresponding to full-cooperation joint decoding bound, and $I_{sel,opt}(x)$ is the DF upper bound optimal power distribution. The DF achievable rates are computed using (50). Note that for single session $C_{coop} = \log(1 + P_r)$, and for multi-session $C_{coop} = P_R$. ($P_s = 20$ dB).

2) **Reliable throughput enhancement (RTE)** - common information streaming of data packets. In this case it is required that both users receive exactly the same information reliably. Hence after the end of the last cooperation session, an ACK signal is returned to the transmitter indicating what the highest common decoded layer was.

The *information enhancement* setting for transmitting common information for two users essentially achieves the **same average rates** as those derived for a single user above.

Our results can be adapted to the *reliable throughput enhancement* setting in the following way. As the maximal decoded layer depends on the actual fading gain, the reliable broadcasting rate is controlled by the smaller equivalent fading after cooperation. This is specified for all cooperation schemes:

- *Naive AF* - the equivalent fading gain $s_b = s_1 + \frac{P_r s_2}{1+P_s s_2+P_r}$ is replaced by $s_b^{RTE} = \min\left\{s_1 + \frac{P_r s_2}{1+P_s s_2+P_r}, s_2 + \frac{P_r s_1}{1+P_s s_1+P_r}\right\}$. Then the corresponding CDF allows com-putation of optimal power allocation and average rates for RTE setting, as in (14).

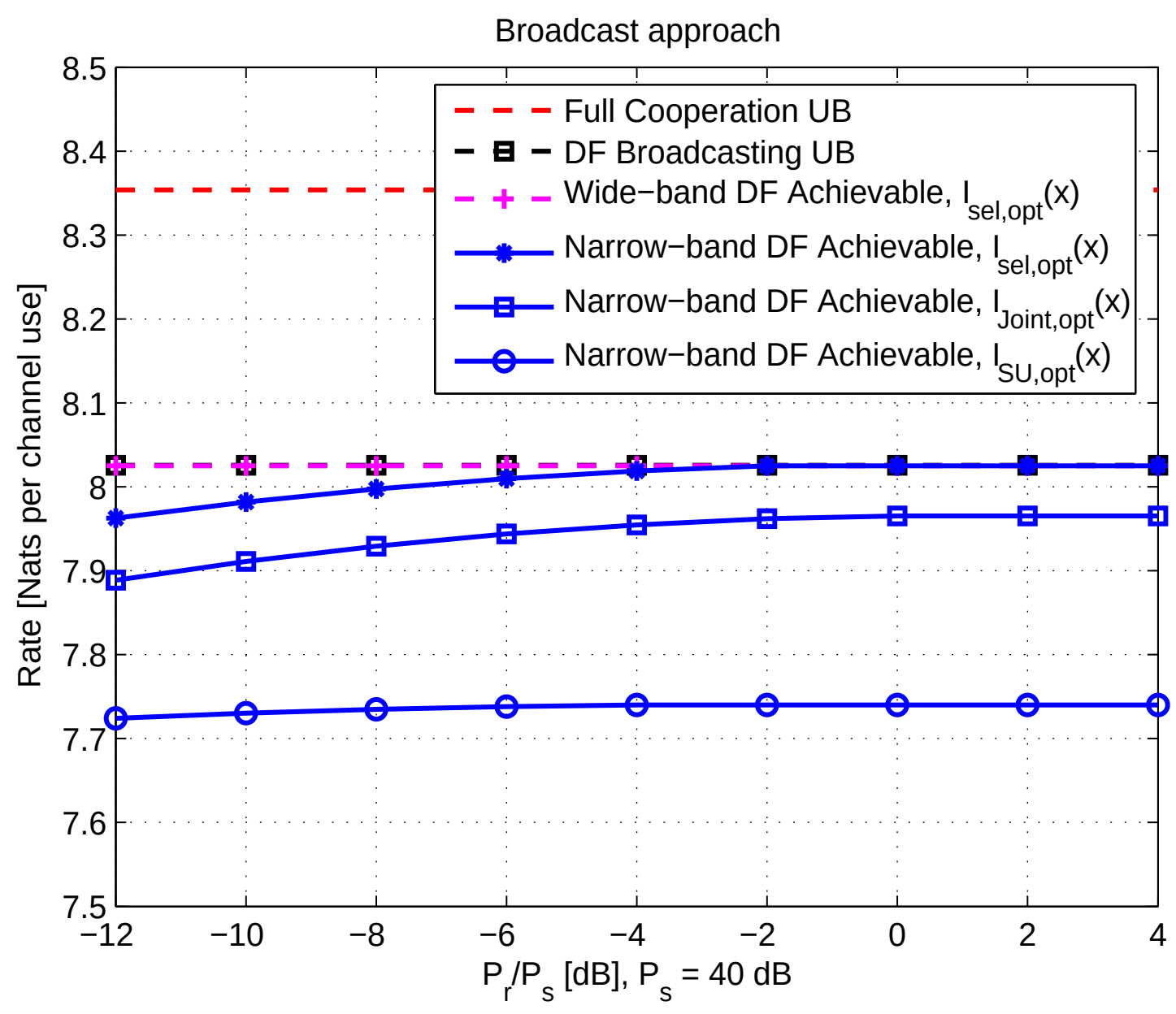

Fig. 10. Broadcast approach: average rates of DF achievable approaches, as function of the channels quality ratio $\frac{P_r}{P_s}$. The approaches are compared as function of the broadcasting power allocation function, where $I_{SU,opt}(x)$ refers to the single user optimal power allocation, and $I_{Joint,opt}(x)$ denotes the function corresponding to full-cooperation joint decoding bound, and $I_{sel,opt}(x)$ is the DF upper bound optimal power distribution. The DF achievable rates are computed using (50). Note that for single session $C_{coop} = \log(1 + P_r)$, and for multi-session $C_{coop} = P_R$. ($P_s = 40$ dB).

- *Separate Preprocessing AF* - the equivalent fading gain $s_a = s_1 + \frac{P_r s_2}{1+s_2\cdot\max(I(s_1),I(s_2))+P_r}$, is replaced by $s_a^{RTE} = \min\left\{s_1 + \frac{P_r s_2}{1+s_2\cdot\max(I(s_1),I(s_2))+P_r}, s_2 + \frac{P_r s_1}{1+s_1\cdot\max(I(s_1),I(s_2))+P_r}\right\}$. Then the corresponding CDF allows computation of optimal power allocation and average rates for RTE setting, as in (25).
- *Multi-session AF with broadband cooperation* - the equivalent fading gain (29) is replaced by $s_b^*$, which is defined there, below (29).
- *Naive CF* - the fading gain $s_{NWZ} = s_2 + \frac{s_1(1+s_2P_s)P_r}{(1+P_r)(1+s_2P_s)+s_1P_s}$ is replaced by $s_{NWZ}^{RTE} = \min\left\{s_1 + \frac{s_2(1+s_1P_s)P_r}{(1+P_r)(1+s_1P_s)+s_2P_s}, s_2 + \frac{s_1(1+s_2P_s)P_r}{(1+P_r)(1+s_2P_s)+s_1P_s}\right\}$, where the CDF of $s_{NWZ}^{RTE}$ allows computation of maximal achievable RTE rates, as in (37)-(38).
- *CF with Separate Preprocessing* - similarly, the equivalent fading gain $s_a^{(1)} = s_1 + \frac{s_2P_r(1+s_1I(s_2))}{(1+P_r)(1+s_1I(s_2))+s_2I(s_2)}$ has to be replaced by $s_a^{(1)} = \min\left\{s_1 + \frac{s_2P_r(1+s_1I(s_2))}{(1+P_r)(1+s_1I(s_2))+s_2I(s_2)}, s_2 + \frac{s_1P_r(1+s_2I(s_1))}{(1+P_r)(1+s_2I(s_1))+s_1I(s_1)}\right\}$.

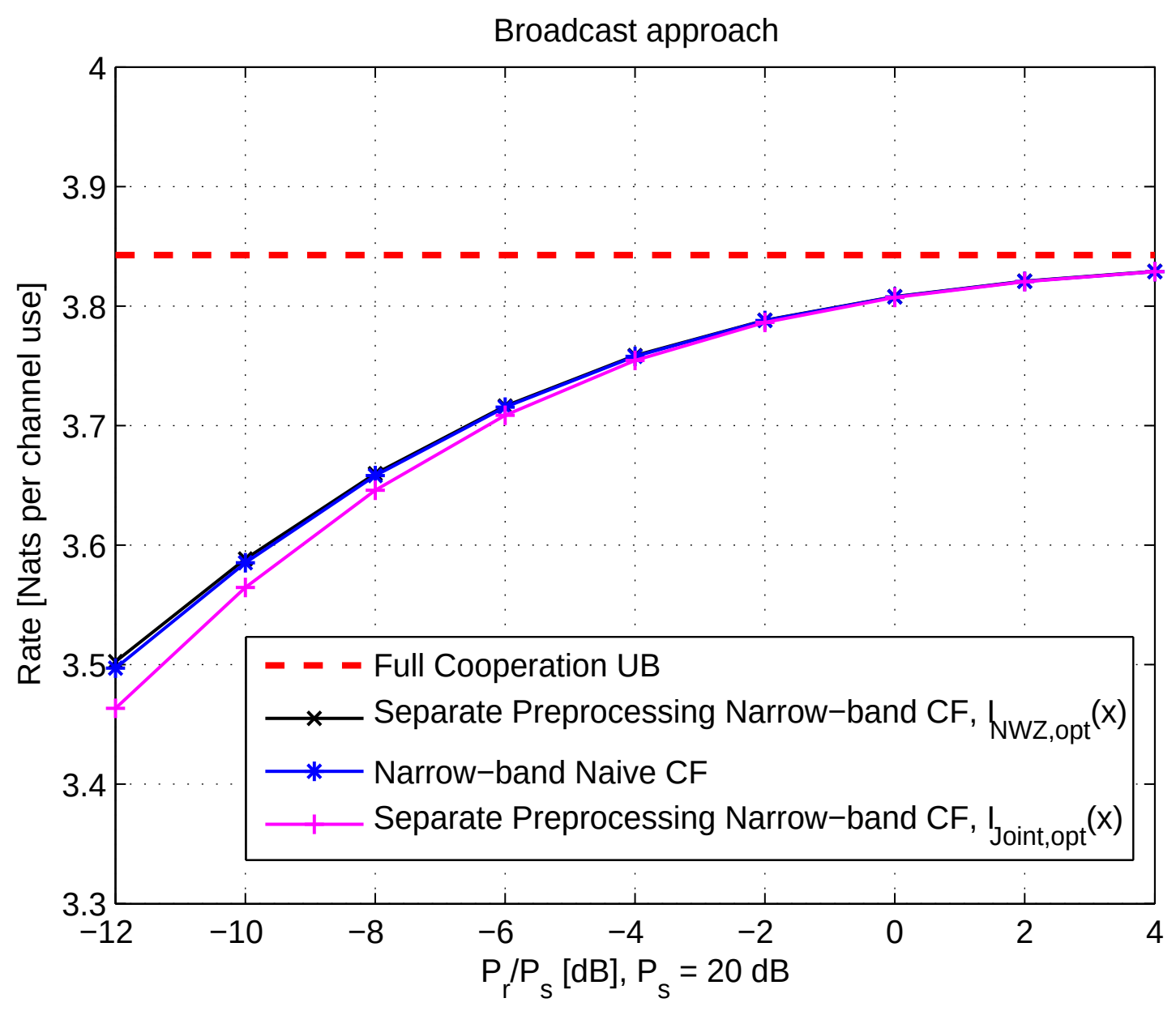

Fig. 11. Broadcast approach: average rates of CF achievable approaches, as function of the channels quality ratio $\frac{P_r}{P_s}$. The approaches are compared as function of the broadcasting power allocation function, where $I_{Joint,opt}(x)$ denotes the power distribution corresponding to full-cooperation joint decoding bound, and $I_{NWZ,opt}(x)$ is the optimal power allocation for naive WZ processing. The CF achievable rates are computed using (40) ($P_s = 20$ dB).

- *Multi-session CF with separate preprocessing* - the equivalent fading gain $s_{ms}$ (45) is replaced by $s_b^{(k)}$, which is defined in (47).

These direct permutations of the equivalent fading gains allow analysis of the RTE setting. This turns the extension of our results for the RTE scheme to be straightforward.

## X. Conclusion

We have considered several cooperation strategies for transmission to colocated users. The original data is intended to one of the users, and in the network setting examined, a colocated user receives another copy of the original signal and cooperates with the destined user to improve decoding at the destination. As the transmitter has no access to CSI, the broadcast approach is used along with the various cooperation strategies. We have examined the naive AF, and its improved version, namely separate preprocessing AF. In the latter, the users decode individually as many layers as they can, subtract the common information and forward a scaled version of the residual signal. In a multi-session AF approach, with a total cooperation power limitation $P_r$, each user tries to decode as many layers as possible using the inputs of the previous cooperation session. Then the users remove common information and scale the residual signal for the next session of cooperation. We give an explicit formulation for large number of sessions, with a fractional power allocation for every session, and an overall power constraint $P_r$ for the whole cooperation duration. This approach may be used in applications where rapidly changing channel does not allow CSI acquisition at the transmitter, and the decoding processing delay constraints are relatively relaxed (due to multi-sesion), and average throughput is to be optimized. When considering average delay as the figure of merit, other broadcasting strategies can be optimal [37]. In this case, multi-session imposes additional delays on the processing per packet.

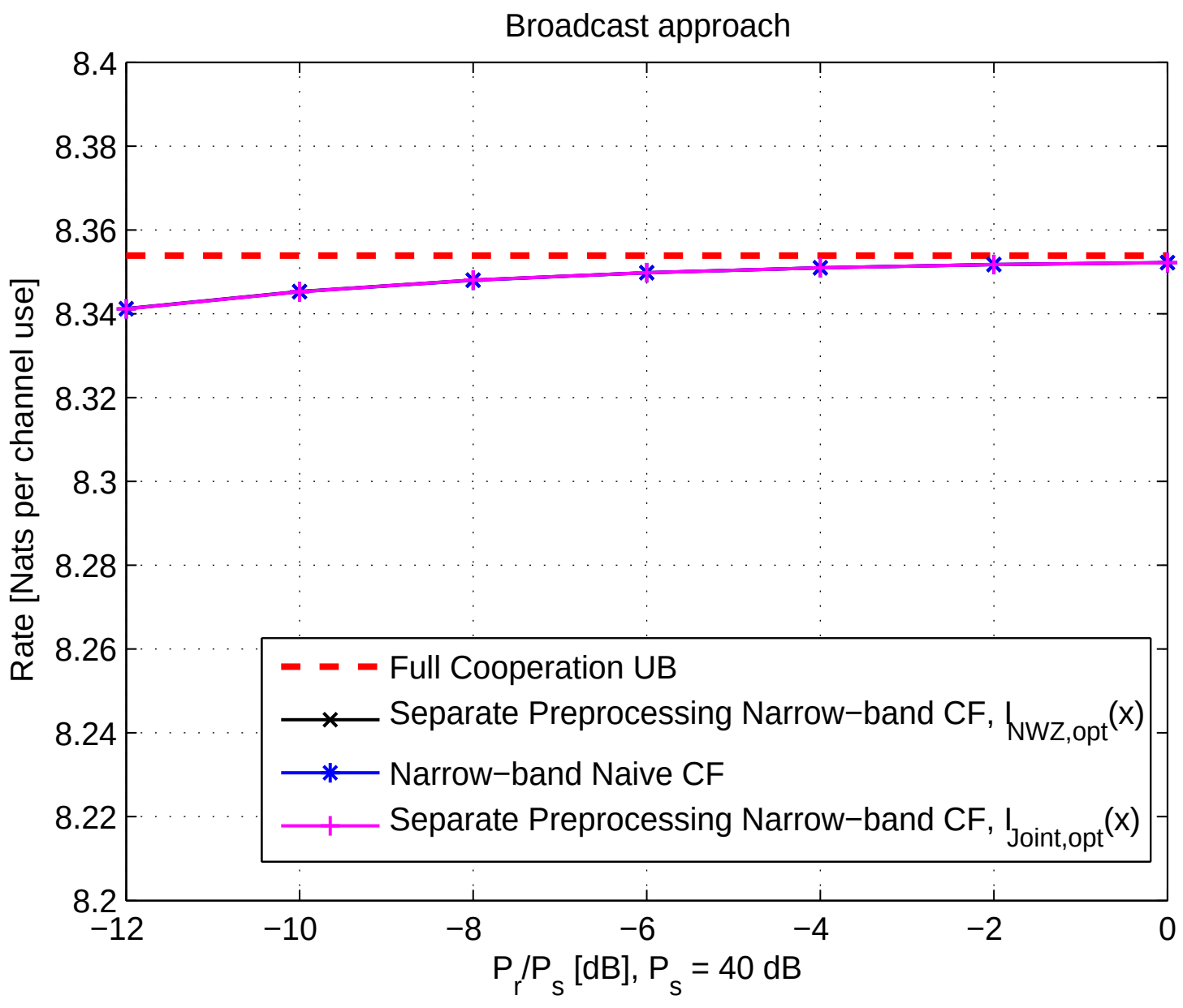

Fig. 12. Broadcast approach: average rates of CF achievable approaches, as function of the channels quality ratio $\frac{P_r}{P_s}$. The approaches are compared as function of the broadcasting power allocation function, where $I_{Joint,opt}(x)$ denotes the power distribution corresponding to full-cooperation joint decoding bound, and $I_{NWZ,opt}(x)$ is the optimal power allocation for naive WZ processing. The CF achievable rates are computed using (40) ($P_s = 40$ dB).

Another cooperation approach considered is CF. In a naive approach the colocated user performs Wyner-Ziv (WZ) compression, and forwards it to the destination user. For the naive CF we derive explicit expressions for the equivalent fading gains, which allows computation of maximal achievable rates. An improved version of this approach for multi-session cooperation, is presented, where for each session, the WZ compression uses all information collected in previous sessions as side information for decoding. This brings notions such as successive refinable WZ coding. Implicit expressions are derived for the equivalent fading gain in a multi-session WZ cooperation.

We consider also DF cooperation with a single session only. That is since DF is unsuitable for multi-session cooperation, as after the first session there is nothing the destination can send back to the other user to improve its decoding.

Numerical results show that narrow-band separate processing CF outperforms all other considered approaches (for which average rates were computed). The narrow-band naive CF already closely approximates the joint decoding broadcasting upper bound. The multi-session AF with a sub-optimal broadcasting power allocation also approximates the broadcasting upper bound in a wide range of SNRs and $P_r/P_s$ ratios. In light of the multi-session gains in the AF technique, and the good performance of the CF, we expect the multi-session CF, to present very good results, even for small SNRs. The DF numerical results show that the DF broadcasting upper bound is achieved by wide-band cooperation, and only marginally degrades for a narrow-band cooperation link.

# APPENDIX A

## AF WITH SEPARATE PREPROCESSING EQUIVALENT FADING

The CDF of $s_a$ can be derived separately for $s_1 > s_2$ and for $s_2 \geq s_1$. This results in the following expression

$$F_{s_a}(x) = \int_0^{\phi_1^{-1}(x)} du f_{s_1}(u) \int_u^{\phi_2(u)} dv f_{s_2}(v) + \int_0^{\phi_1^{-1}(x)} du f_{s_2}(u) \int_u^{\phi_3(u)} dv f_{s_1}(v) \tag{A.1}$$

where

$$\begin{aligned} \phi_1(u) &= u + \tfrac{uP_r}{1+uI(u)+P_r} \\ \phi_2(u) &= \max\left(u, x - \tfrac{uP_r}{1+uI(u)+P_r}\right) \\ \phi_3(u) &= \max\left(u, \phi_4(x-u)\right) \end{aligned} \tag{A.2}$$

where

$$\phi_4(x-u) = \begin{cases} \frac{(1+P_r)(x-u)}{P_r - I(u)(x-u)} & P_r - I(u)(x-u) > 0 \\ \infty & P_r - I(u)(x-u) <= 0 \end{cases} \tag{A.3}$$

For a Rayleigh fading channel, the CDF in (A.1) reduces to a single integral expression, as specified in (26).

## APPENDIX B
### MULTI-SESSION AF WITH SEPARATE PREPROCESSING EQUIVALENT FADING

We assume without loss of generality that $s_1 > s_2$, and that both users are aware of the index $s_2$. Thus they decode independently all layers up to $s_2$ and remove the decoded signal from the received channel output. Then, they exchange the residual signal, amplified to power $\delta_1$. User $i = 1$ can now decode up to the layer associated with the equivalent fading, similar to (24), but with different AF power, namely

$$s_a^{(1)} = s_1 + \frac{\delta_1 s_2}{1 + s_2 I(s_2) + \delta_1} \tag{B.1}$$

where the superscript of $s_a$ indicates the cooperation session index. The received signal of the first session, at $i = 1$, is given by

$$\boldsymbol{y}_{1,2}^{(1)} = \sqrt{\beta_b^{(1)}}(h_2 \boldsymbol{x}_I^{(1)} + \boldsymbol{n}_2) + \boldsymbol{n}_c^{(1)}, \tag{B.2}$$

where $n_c^{(1)}$ is the noise on the cooperation link, and $\beta_b^{(1)} = \frac{\delta_1}{1+I(s_2)s_2}$. Where for user $i = 2$ similar relations exist. In the second session there is a higher common decoded layer $s_b^{(2)} \geq s_2$, and since both users forwarded with the same power, $s_a^{(2)} \geq s_b^{(2)}$. Thus both users remove decoded layers up to $s_b^{(2)}$ and amplify the residual signal over to the other user. The received signal at the $k^{th}$ session, at user $i = 1$ is given by

$$\boldsymbol{y}_{1,2}^{(k)} = \sqrt{\beta_b^{(k)}}(h_2 \boldsymbol{x}_I^{(k)} + \boldsymbol{n}_2) + \boldsymbol{n}_c^{(k)}, \tag{B.3}$$

where similar expression exists for user $i = 2$. In order to perform optimal decoding, the following is done. All decoded layers are cancelled out from the cooperation inputs, $\{\boldsymbol{y}_{1,2}^{(i)}\}_{i=1}^{k-1}$, for user $i = 1$, and a maximal ratio combining of all inputs is performed. Thus the equivalent SNR for decoding at the $k^{th}$ session in user $i = 1$ is

$$s_a^{(k)} \triangleq s_1 + \frac{s_2 \sum\limits_{i=1}^{k} \beta_b^{(i)}}{1 + \sum\limits_{i=1}^{k} \beta_b^{(i)}}, \tag{B.4}$$

where

$$\beta_b^{(i)} = \frac{\delta_i}{1 + I\left(s_b^{(i)}\right) s_2}. \tag{B.5}$$

The common layer, decoded at the $k^{th}$ session, is associated with $s_b^{(k)}$, which is equal to:

$$s_b^{(k)} \triangleq s_2 + \frac{s_1 \sum\limits_{i=1}^{k} \beta_a^{(i)}}{1 + \sum\limits_{i=1}^{k} \beta_a^{(i)}}. \tag{B.6}$$

where

$$\beta_a^{(i)} = \frac{\delta_i}{1 + I\left(s_b^{(i)}\right) s_1}. \tag{B.7}$$

Note that also $s_b^{(k)} \leq s_a^{(k)}$, so that $i = 1$ will decode all that is decoded by $i = 2$, which is true since they both forward through $y_{1,2}$ and $y_{2,1}$, with the same powers: $\{\delta_i\}_1^k$, and since $s_1 \geq s_2$. The equivalent fading after the $k^{th}$ session at user $i = 1$ and $i = 2$ is explicitly given by using (B.4) and (B.5),

$$s_a^{(k)} = s_1 + s_2 \frac{\sum\limits_{i=1}^{k} \frac{\delta_i}{1+I\left(s_b^{(i)}\right)s_2}}{1 + \sum\limits_{i=1}^{k} \frac{\delta_i}{1+I\left(s_b^{(i)}\right)s_2}} \tag{B.8}$$

and respectively

$$s_b^{(k)} = s_2 + s_1 \frac{\sum\limits_{i=1}^{k} \frac{\delta_i}{1+I\left(s_b^{(i)}\right)s_1}}{1 + \sum\limits_{i=1}^{k} \frac{\delta_i}{1+I\left(s_b^{(i)}\right)s_1}}. \tag{B.9}$$

Since $I(s)$ is a decreasing function of $s$, $\{s_a^{(k)}\}$ is monotonically increasing, upper bounded by $s_1 + s_2$, and thus it also converges to a limit $s_a^*$ which is upper bounded by $s_u$, given implicitly as

$$s_u = s_1 + \frac{P_r s_2}{1 + s_2 I(s_u) + P_r}. \tag{B.10}$$

Let us focus on the case of infinitely many sessions, each with some infinitely small power $\{\delta_i\}_{i=1}^{\infty}$. First, from equation (B.8) we can write

$$\Delta_b^k \triangleq s_b^{(k)} - s_b^{(k-1)} = \frac{s_1 \delta_k}{(1 + X_{k-1})\left[\delta_k + \left(1 + s_1 I\left(s_b^{(k)}\right)\right)(1 + X_{k-1})\right]} \tag{B.11}$$

where

$$X_{k-1} \triangleq \sum_{i=1}^{k-1} \frac{\delta_i}{1 + s_1 I\left(s_b^{(i)}\right)}. \tag{B.12}$$

For infinitely many sessions $k \to \infty$, $\delta_k \to 0$, while $\sum_1^k \delta_i = P_r$.

$$\begin{aligned} \tfrac{\delta_k}{\Delta_b^k} &= \tfrac{1}{s_1}(1 + X_{k-1}) \left[\delta_k + \left(1 + s_1 I\left(s_b^{(k)}\right)\right)(1 + X_{k-1})\right] \\ &= \tfrac{1}{s_1}(1 + X_{k-1}) \left[\delta_k + \left(1 + s_1 I\left(s_b^{(k-1)} + \Delta_b^k\right)\right)(1 + X_{k-1})\right] \end{aligned} \tag{B.13}$$

Taking the limit of (B.13) results with

$$\lim_{\Delta_b^k \to 0} \frac{\delta_k}{\Delta_b^k} = \delta'(s_b) = \frac{1}{s_1}(1 + s_1 I(s_b))(1 + X(s_b))^2 \triangleq \rho(s_b), \tag{B.14}$$

where

$$X(s) = \lim_{\Delta_b \to 0, k \to \infty} X_{k-1} = \lim_{\Delta_b \to 0} \sum_{j=1}^{k-1} \frac{\delta_j}{\Delta_b} \frac{\Delta_b}{1 + s_1 I(s_2 + j\Delta_b)} = \int_{s_2}^{s} \frac{\rho(\sigma) d\sigma}{1 + s_1 I(\sigma)}. \tag{B.15}$$

where we have assumed that $\Delta_b = \Delta_b^k \; \forall k$, which means that $\delta_k$ is chosen every session according to $\frac{d\delta(s)}{ds}$. Rewriting (B.14) gives

$$\frac{\rho(s)}{1 + s_1 I(s)} = \frac{1}{s_1}(1 + X(s))^2, \tag{B.16}$$

where the left hand side is the integrand in (B.15). Hence the following equality holds

$$X'(s) = \frac{1}{s_1}(1 + X(s))^2, \tag{B.17}$$

which can be solved, using the initial condition $X(s_2) = 0$,

$$X(s) = \frac{s - s_2}{s_1 + s_2 - s}. \tag{B.18}$$

This means that

$$\rho(s_b) = (1 + s_1 I(s_b)) \frac{s_1}{(s_1 + s_2 - s_b)^2}, \tag{B.19}$$

where using $\int_{s_2}^{s_b^*} \rho(s) ds = P_r$, we get the following implicit equation from which we can get the resulting $s_b^*$, which corresponds to the channel available at user $i = 2$ after infinitely many conference sessions, with a total power of $P_r$,

$$\int_{s_2}^{s_b^*} \frac{s_1}{(s_1 + s_2 - \sigma)^2} [1 + s_1 I(\sigma)] d\sigma = P_r. \tag{B.20}$$

The equivalent SNR $s_a^*$ of user $i = 1$ is more interesting, since it can decode more layers. From the above definition of $\beta_a^{(i)}$ (B.7), in the limit of $\Delta_b \to 0$,

$$Z(s) = \lim_{k\to\infty} \sum_{i=1}^{k} \beta_a^{(i)} = \int_{s_2}^{s} \frac{\rho(\sigma)}{1 + s_2 I(\sigma)} d\sigma \tag{B.21}$$

and using the result of (B.19), we get an implicit expression for $Z(s)$,

$$Z(s) = \int_{s_2}^{s} \frac{1 + s_1 I(\sigma)}{(1 + s_2 I(\sigma))} \frac{s_1}{(s_1 + s_2 - \sigma)} d\sigma \tag{B.22}$$

Once we have solved $s_b^*$ from (B.20), we can find $s_a^*$ by

$$s_a^* = s_1 + s_2 \frac{Z(s_b^*)}{1 + Z(s_b^*)}. \tag{B.23}$$

Note that due to the assumption that $s_1 \geq s_2$, the destination user can decode up to $s_a^*$. Clearly, for $s_1 < s_2$, the destination user will be able to decode only up to $s_b^*$. ■

## APPENDIX C

### Naive Compress and Forward Equivalent Fading

The mutual information expressions are directly derived from (33)-(35)

$$R_{WZ,2}(h_1, h_2) = I(x_s; y_2, \hat{y}_1 | h_1, h_2) = \log\left(1 + s_2 P_s + \frac{s_1 P_s}{1 + \sigma^2}\right), \tag{C.1}$$

where $s_i = |h_i|^2$.

Let us evaluate $\sigma$ from (C.1):

$$\begin{aligned} I(y_1; \hat{y}_1 | h_1) &= \log(1 + \tfrac{s_1 P_s + 1}{\sigma^2}) \\ I(y_2; \hat{y}_1 | h_1, h_2) &= \log\left(1 + \tfrac{s_1 s_2 P_s^2}{(1+s_2 P_s)(1+\sigma^2+s_1 P_s) - s_1 s_2 P_s^2}\right). \end{aligned} \tag{C.2}$$

The derivation of $I(y_2; \hat{y}_1 | h_1, h_2)$ follows from its definition, $I(y_2; \hat{y}_1 | h_1, h_2) = h(\hat{y}_1 | h_1) + h(y_2 | h_2) - h(\hat{y}_1, y_2 | h_1, h_2)$. It follows immediately that

$$\begin{aligned} h(\hat{y}_1 | h_1) &= \log \pi e (1 + \sigma^2 + s_1 P_s) \\ h(y_2 | h_1) &= \log \pi (1 + s_2 P_s) \end{aligned} \tag{C.3}$$

The covariance matrix of $(\hat{y}_1, y_2)$ is given by

$$\Lambda_{\hat{y}_1, y_2} = E \begin{bmatrix} y_2 \\ \hat{y}_1 \end{bmatrix} [y_2^* \ \hat{y}_1^*] = \begin{bmatrix} 1 + s_2 P_s & h_1^* h_2 P_s \\ h_1 h_2^* P_s & 1 + \sigma^2 + s_1 P_s \end{bmatrix} \tag{C.4}$$

From the covariance matrix, the entropy of $(\hat{y}_1, y_2)$ can be computed

$$h(\hat{y}_1, y_2 | h_1, h_2) = \log \det(\pi e \Lambda_{\hat{y}_1, y_2}) = \log \pi^2 e^2 \left((1 + s_2 P_s)(1 + \sigma^2 + s_1 P_s) - s_1 s_2 P_s^2\right) \tag{C.5}$$

Combining (C.3), and (C.5) gives $I(y_2;\hat{y}_1)$ in (C.2). Next we note that the Wyner-Ziv compression rate is

$$I(y_1;\hat{y}_1|h_1) - I(y_2;\hat{y}_1|h_1,h_2) = \log\frac{(1+\sigma^2+s_1P_s)(s_1P_s+(1+\sigma^2)(1+s_2P_s))}{\sigma^2\left[(1+\sigma^2)(1+s_2P_s)+s_1P_s+s_1s_2P_s^2\right]} \qquad \text{(C.6)}$$

The capacity of the cooperation channel $C_{coop}$ restricts the compression rate, according to the condition in (34). When performing compression as function of $(s_1,s_2)$, such that for every such pair there is a different codebook the best compression is achieved when condition (34) is satisfied with equality. This means that for the narrow-band link with $C_{coop}=\log(1+P_r)$,

$$\sigma^2_{NB} = \frac{1+s_1P_s+s_2P_s}{P_r(1+s_2P_s)}, \qquad \text{(C.7)}$$

and for the wide-band transmission with $C_{coop}=P_r$,

$$\sigma^2_{WB} = \frac{1+s_1P_s+s_2P_s}{(e^{P_r}-1)(1+s_2P_s))} \qquad \text{(C.8)}$$

To summarize the results, the achievable rate, governed by the cooperation channel capacity and fading gains for $C_{coop}=\log(1+P_r)$, is given by

$$I(x_s;y_2,\hat{y}_1)_{NB} = \log\left(1+s_2P_s+\frac{s_1P_s(1+s_2P_s)P_r}{(1+P_r)(1+s_2P_s)+s_1P_s}\right). \qquad \text{(C.9)}$$

We continue with analysis of narrow-band cooperation only, as the same results can be directly obtained for the wide-band cooperation link. It may be noticed that the higher $P_r$ is the closer the performance can get to coherent combining (SIMO processing).

Let us calculate the average rate which is decoded at each user, using the Wyner-Ziv compression. First we need to calculate $F_s(u)$ which is

$$F_s(u) = \text{Prob}(s\le u) = \text{Prob}\left(s_2+\frac{s_1(1+s_2P_s)P_r}{(1+P_r)(1+s_2P_s)+s_1P_s}\le u\right) = \int f_{s_2}(v)\text{Prob}(s\le u|s_2=v)dv. \qquad \text{(C.10)}$$

This integral can be written as:

$$F_s(u) = \int_{\max\left\{\frac{P_su-P_r}{P_s(1+P_r)},0\right\}}^{u} dv f_{s_2}(v)\int_0^{\frac{(u-v)(1+P_r)(1+vP_s)}{(1+vP_s)P_r-(u-v)P_s}} du f_{s_1}(u) + \int_0^{\max\left\{\frac{P_su-P_r}{P_s(1+P_r)},0\right\}} dv f_{s_2}(v) =$$

$$1-e^{-u}-\int_{\max\left\{\frac{P_su-P_r}{P_s(1+P_r)},0\right\}}^{u} dv\exp\left(-v-\frac{(u-v)(1+P_r)(1+vP_s)}{(1+vP_s)P_r-(u-v)P_s}\right). \qquad \text{(C.11)}$$

Using the equality[1]:

$$\int dve^{-v-\frac{(u-v)(1+P_r)(1+vP_s)}{(1+vP_s)P_r-(u-v)P_s}} = e^{-\frac{P_r(1+uP_s)^2}{(1+P_r)P_s(v(1+P_r)P_s+P_r-uP_s)}-\frac{uP_rP_s-1}{(P_r+1)P_s}}\left(v+\frac{P_r-uP_s}{(P_r+1)P_s}\right)$$
$$-P_r(1+uP_s)^2((P_r+1)P_s)^{-2}e^{-\frac{uP_rP_s-1}{(P_r+1)P_s}}\mathrm{Ei}\left(1,\frac{P_r(1+uP_s)^2}{(1+P_r)P_s(v(1+P_r)P_s+P_r-uP_s)}\right) \quad \text{(C.12)}$$

we find $F_s(u)$, when $u \geq \frac{P_r}{P_s}$:

$$F_s(u) \;=\; 1-e^{-u}\left(1+\frac{P_r(uP_s+1)}{P_s(P_r+1)}\right)+\frac{P_r(uP_s+1)^2}{P_s^2(P_r+1)^2}e^{-\frac{uP_rP_s-1}{P_s(P_r+1)}}\mathrm{Ei}\left(1,\frac{uP_s+1}{P_s(P_r+1)}\right) \quad \text{(C.13)}$$

and when $u < \frac{P_r}{P_s}$:

$$F_s(u)=1-e^{-u}\left(1+\frac{P_r(uP_s+1)}{P_s(P_r+1)}\right)-\frac{P_r-uP_s}{(1+P_r)P_s}e^{-\frac{u}{P_r-uP_s}}+\frac{P_r(uP_s+1)^2}{P_s^2(P_r+1)^2}e^{-\frac{uP_rP_s-1}{P_s(P_r+1)}}$$
$$\left(\mathrm{Ei}\left(1,\frac{uP_s+1}{P_s(P_r+1)}\right)-\mathrm{Ei}\left(1,\frac{P_r(uP_s+1)^2}{P_s(P_r+1)(P_r-uP_s)}\right)\right). \quad \text{(C.14)}$$

# APPENDIX D

## MULTI-SESSION COMPRESS AND FORWARD EQUIVALENT FADING

Since $I(y_1;\hat{y}_1^{(k)}|\hat{y}_1^{(k-1)},\ldots,\hat{y}_1^{(1)},y_2,x_{s,D}^{(k-1)},h_1,h_2) = I(y_1;\hat{y}_1^{(k)}|\hat{y}_1^{(k-1)},y_2,x_{s,D}^{(k-1)},h_1,h_2)$, the required colocation bandwidth for the $k$-th session is

$$\log(1+\delta_1^{(k)}) = I(y_1;\hat{y}_1^{(k)}|\hat{y}_1^{(k-1)},y_2,x_{s,D}^{(k-1)},h_1,h_2)$$
$$= I(y_1;\hat{y}_1^{(k)}|x_{s,D}^{(k-1)},h_1,h_2)+I(y_2;\hat{y}_1^{(k-1)}|x_{s,D}^{(k-1)},h_1,h_2)-I(y_1;\hat{y}_1^{(k-1)}|x_{s,D}^{(k-1)},h_1,h_2)-I(y_2;\hat{y}_1^{(k)}|x_{s,D}^{(k-1)},h_1,h_2)$$
$$= R_{WZ,C,1}^{(k)} - R_{WZ,C,1}^{(k-1)} \quad \text{(D.1)}$$

(where $R_{WZ,C,i} \triangleq I(y_i;\hat{y}_i^{(k)}|x_{s,D}^{(k-1)},h_1,h_2) - I(y_{3-i};\hat{y}_i^{(k)}|x_{s,D}^{(k-1)},h_1,h_2),\ i=1,2$) and

$$\log(1+\delta_2^{(k)}) = I(y_2;\hat{y}_2^{(k)}|\hat{y}_2^{(k-1)},y_1,x_{s,D}^{(k-1)},h_1,h_2)$$
$$= I(y_2;\hat{y}_2^{(k)}|x_{s,D}^{(k-1)},h_1,h_2)+I(y_1;\hat{y}_2^{(k-1)}|x_{s,D}^{(k-1)},h_1,h_2)-I(y_2;\hat{y}_2^{(k-1)}|x_{s,D}^{(k-1)},h_1,h_2)-I(y_1;\hat{y}_2^{(k)}|x_{s,D}^{(k-1)},h_1,h_2)$$
$$= R_{WZ,C,2}^{(k)} - R_{WZ,C,2}^{(k-1)} \quad \text{(D.2)}$$

where $R_{WZ}$ was defined in (34), and $x_{s,D}^{(k)}$ is the decoded signal in the $k^{th}$ session, by both users. We notice that this can be further improved by using the extra side information at the stronger user, however, for the sake of brevity, we consider only the commonly decoded layers $x_{s,D}^{(k)}$. Considering that we deal with Gaussian channel, if $\{z_{c,j}^i\}_{j=1,i=1}^{2,k-1}$ are independent Gaussian variables with zero mean and variances of $\left\{\left(\sigma_j^{(i-1)}\right)^2 - \left(\sigma_j^{(i)}\right)^2\right\}$, where $\mathrm{E}\left|n_{c,j}^{(i)}\right|^2 = \left(\sigma_j^{(i)}\right)^2$, then

$$j = 1, 2: \; n_{c,j}^{(i)} = y_j + n_{c,j}^{(k)} + \sum_{l=i}^{k-1} z_{c,j}^{(l)}. \tag{D.3}$$

[1] $\mathrm{Ei}(n,x)=\int_1^\infty \frac{e^{-xt}}{t^n}dt$ is the exponential integral

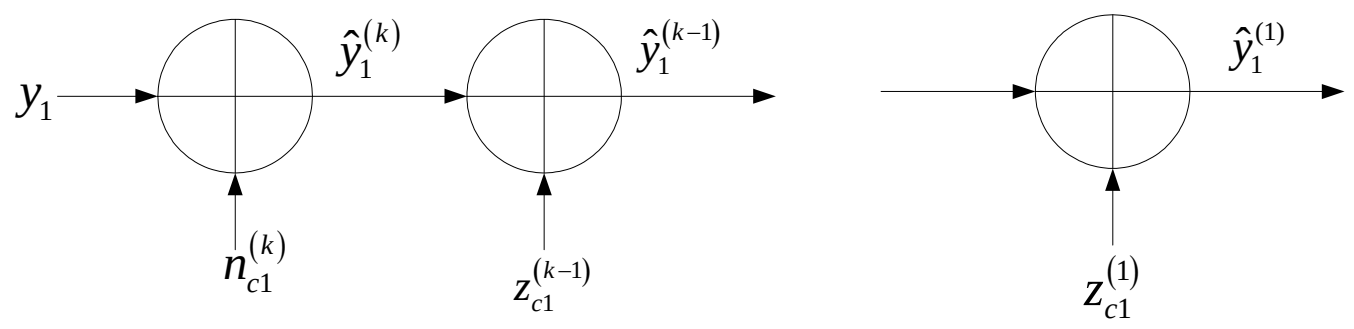

Fig. 13. The Markov relation between the compressed signals

This is seen in figure 13. The compression quality in each session can be recursively calculated by ($\sigma_j^{(0)} = \infty$):

$$\frac{F(\sigma_j^{(k)}, k, s_j, s_{3-j}, s^{(k-1)})}{F(\sigma_j^{(k-1)}, k, s_j, s_{3-j}, s^{(k-1)})} = 1 + \delta_j^{(k)} \tag{D.4}$$

where $s^{(k-1)} = \min\{s_b^{(k-1)}, s_a^{(k-1)}\}$ ($s_b^{(k-1)}, s_a^{(k-1)}$ are defined according to (D.7) and (D.8)), and

$$F(\varsigma, k, s_j, s_{3-j}, s^{(k-1)}) \triangleq \frac{s_j I(s^{(k-1)}) + (1+\varsigma^2)(1 + s_{3-j} I(s^{(k-1)}))}{\varsigma^2(1 + s_{3-j} I(s^{(k-1)}))}. \tag{D.5}$$

Notice that when $k = 1$, (D.4) is indeed identical to the case of single session cooperation, given in equation (C.6). Solving (D.4) for $\sigma_j^{(k)}$ results with:

$$\left(\sigma_j^{(k)}\right)^2 = \left(\sigma_j^{(k-1)}\right)^2 \frac{1 + s_j I(s^{(k-1)}) + s_{3-j} I(s^{(k-1)})}{(1 + s_{3-j} I(s^{(k-1)}))\left[1 + \delta_j^{(k)}\left(1 + \left(\sigma_j^{(k-1)}\right)^2\right)\right] + s_j I(s^{(k-1)})(1 + \delta_j^{(k)})} \tag{D.6}$$

The achievable rate remains (C.1), which are calculated now by

$$s_a^{(k)} = s_1 + \frac{s_2}{1 + (\sigma_2^{(k)})^2} \tag{D.7}$$

$$s_b^{(k)} = s_2 + \frac{s_1}{1 + (\sigma_1^{(k)})^2}, \tag{D.8}$$

and

$$R_{WZ,1}^{(k)} = \log(1 + s_a^{(k)} P_s) \tag{D.9}$$

$$R_{WZ,2}^{(k)} = \log(1 + s_b^{(k)} P_s). \tag{D.10}$$

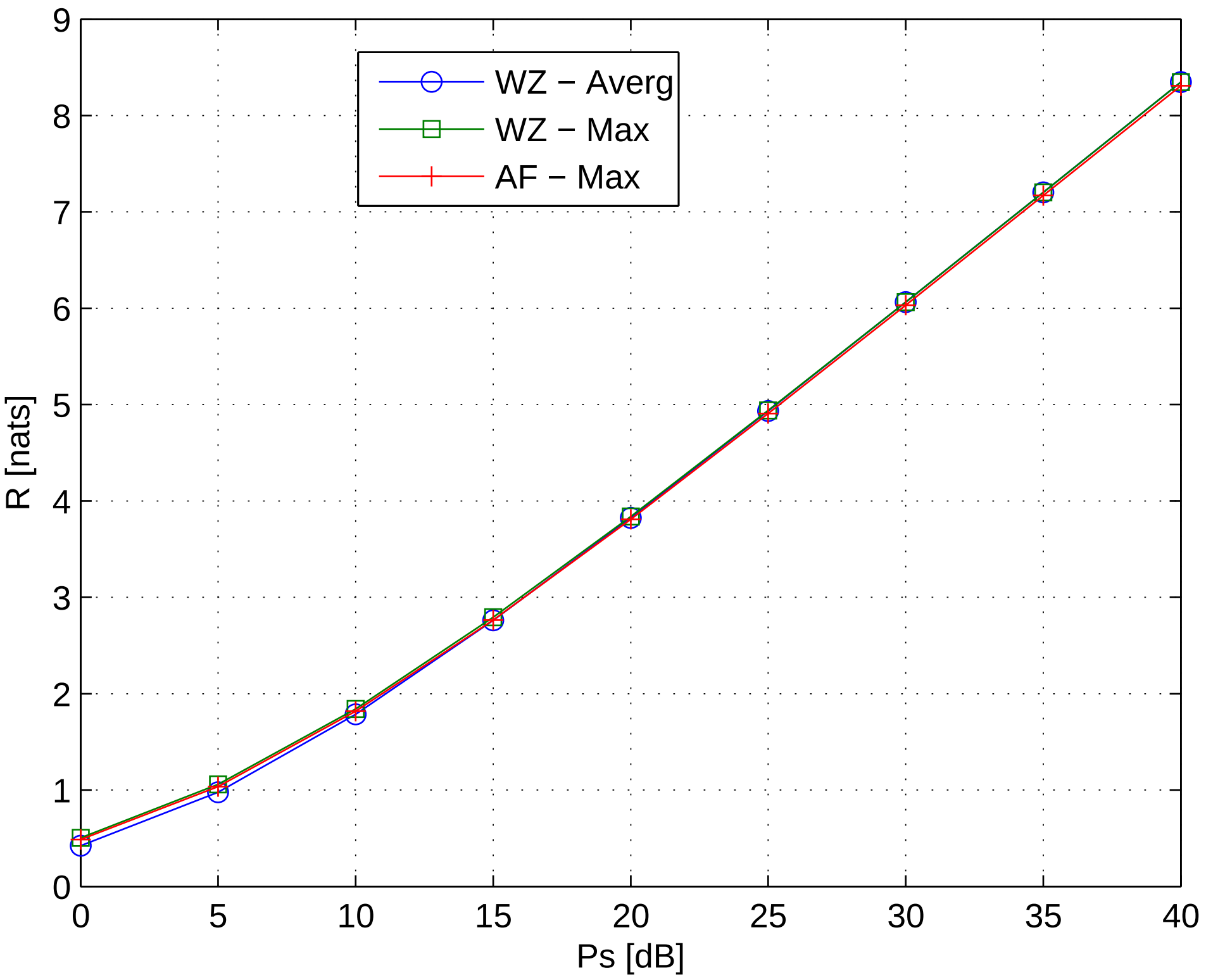

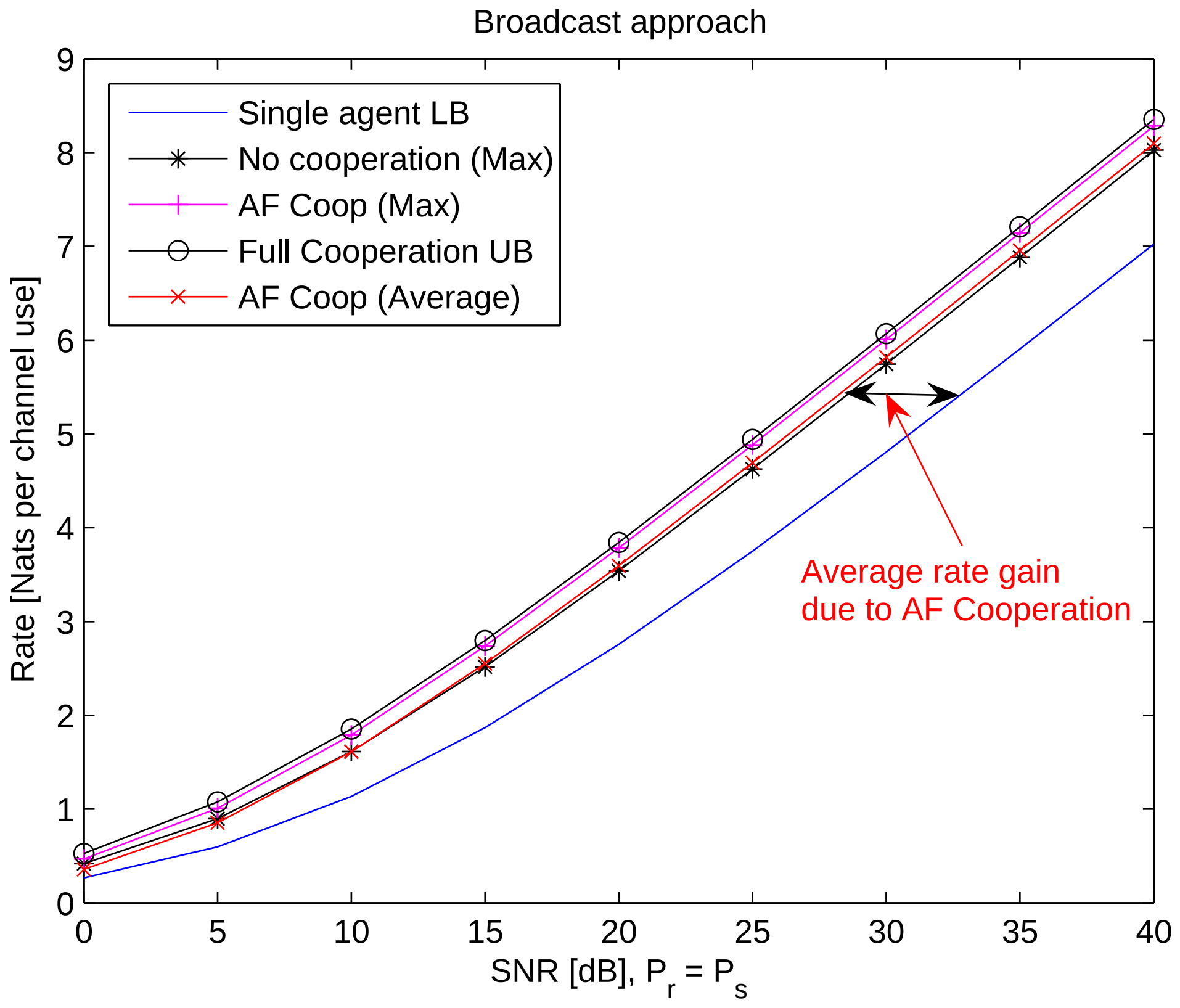

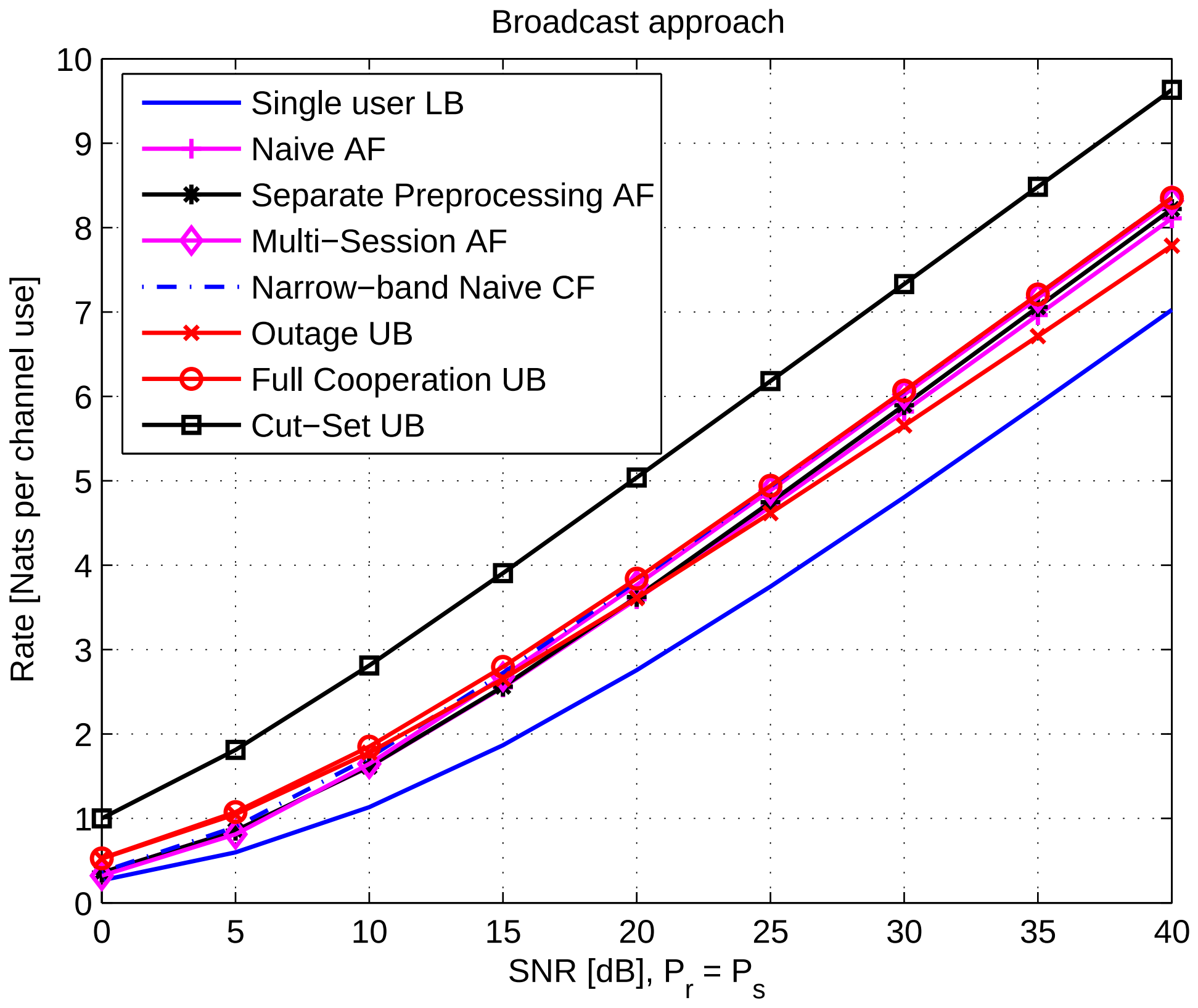

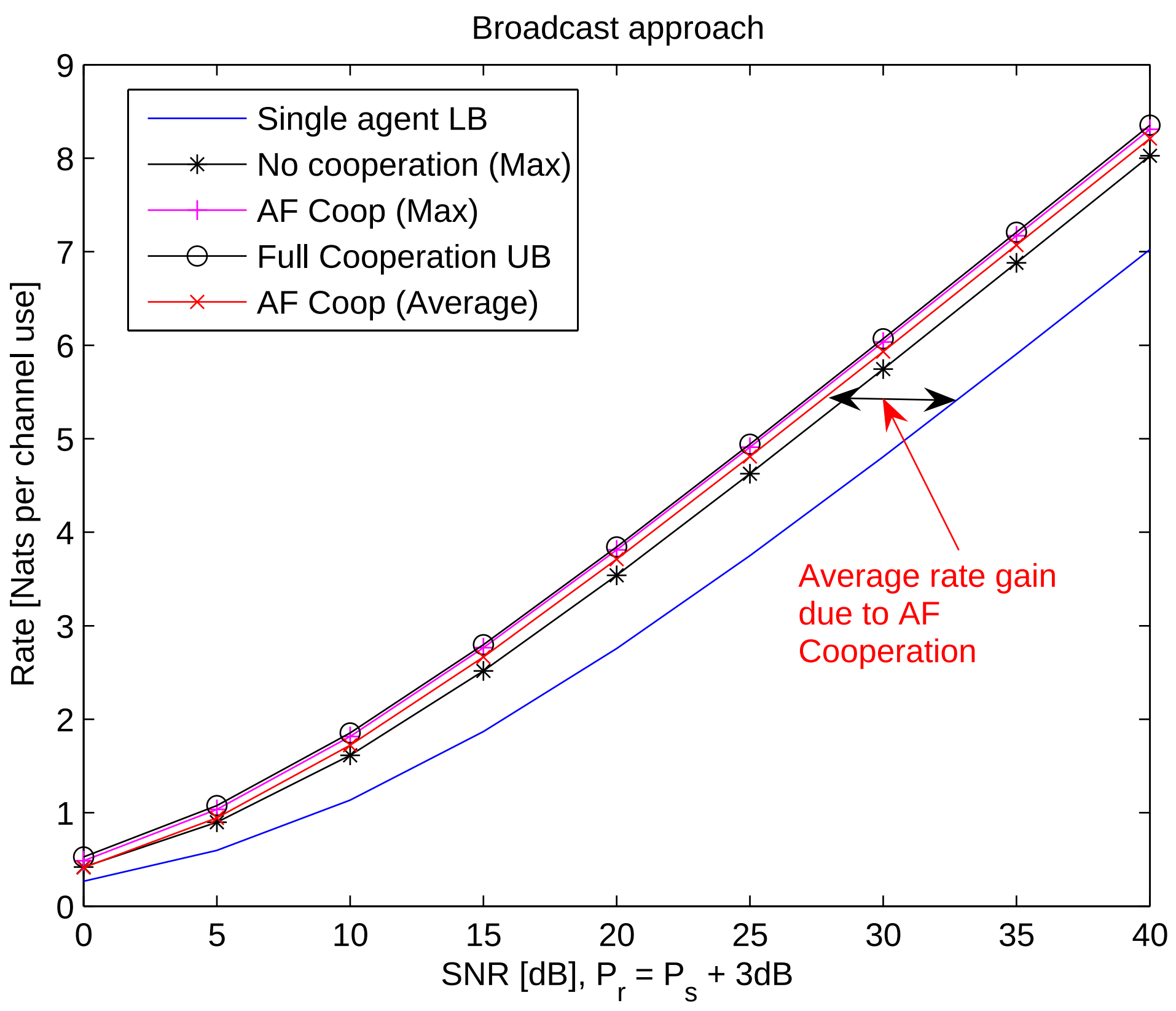

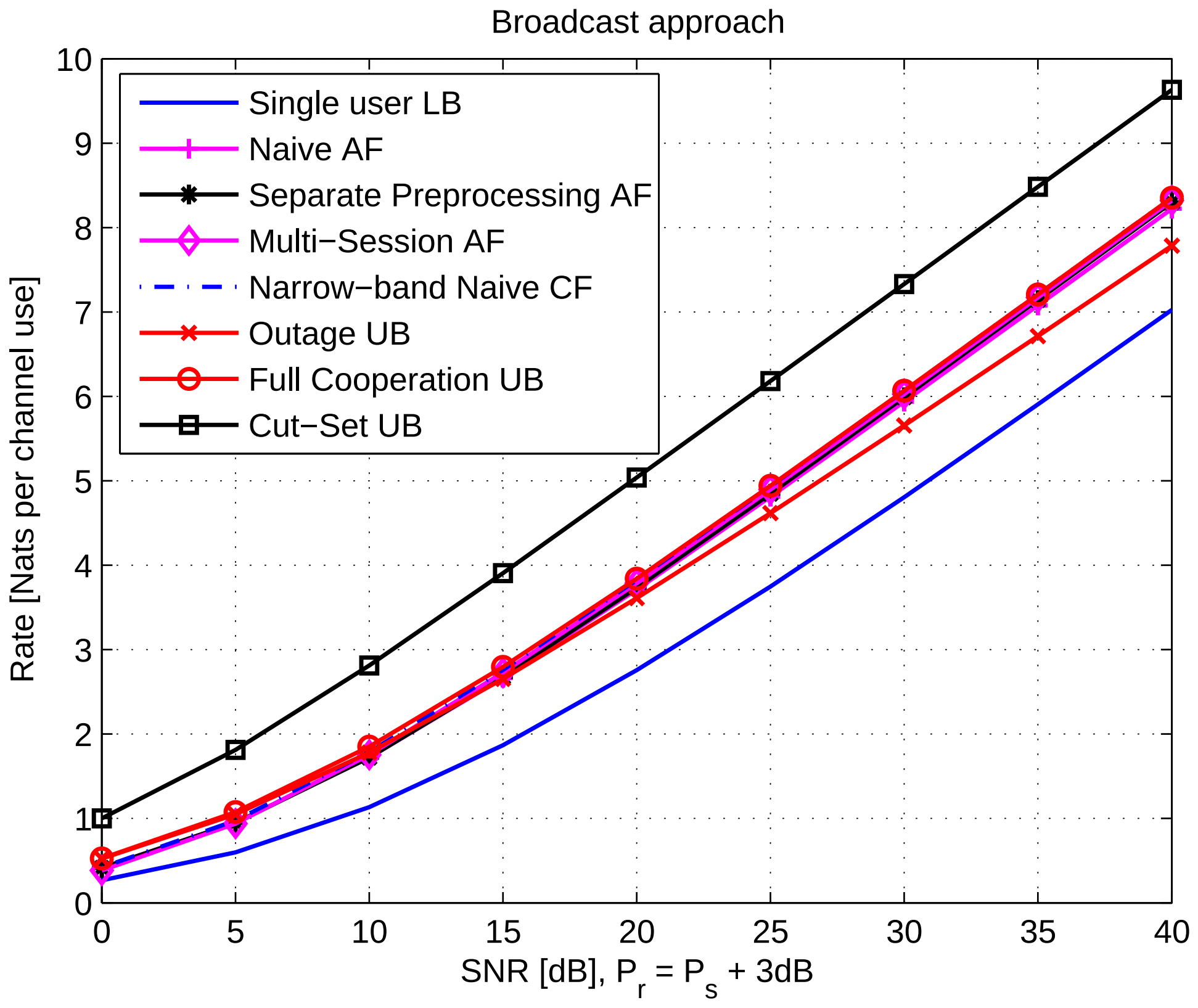

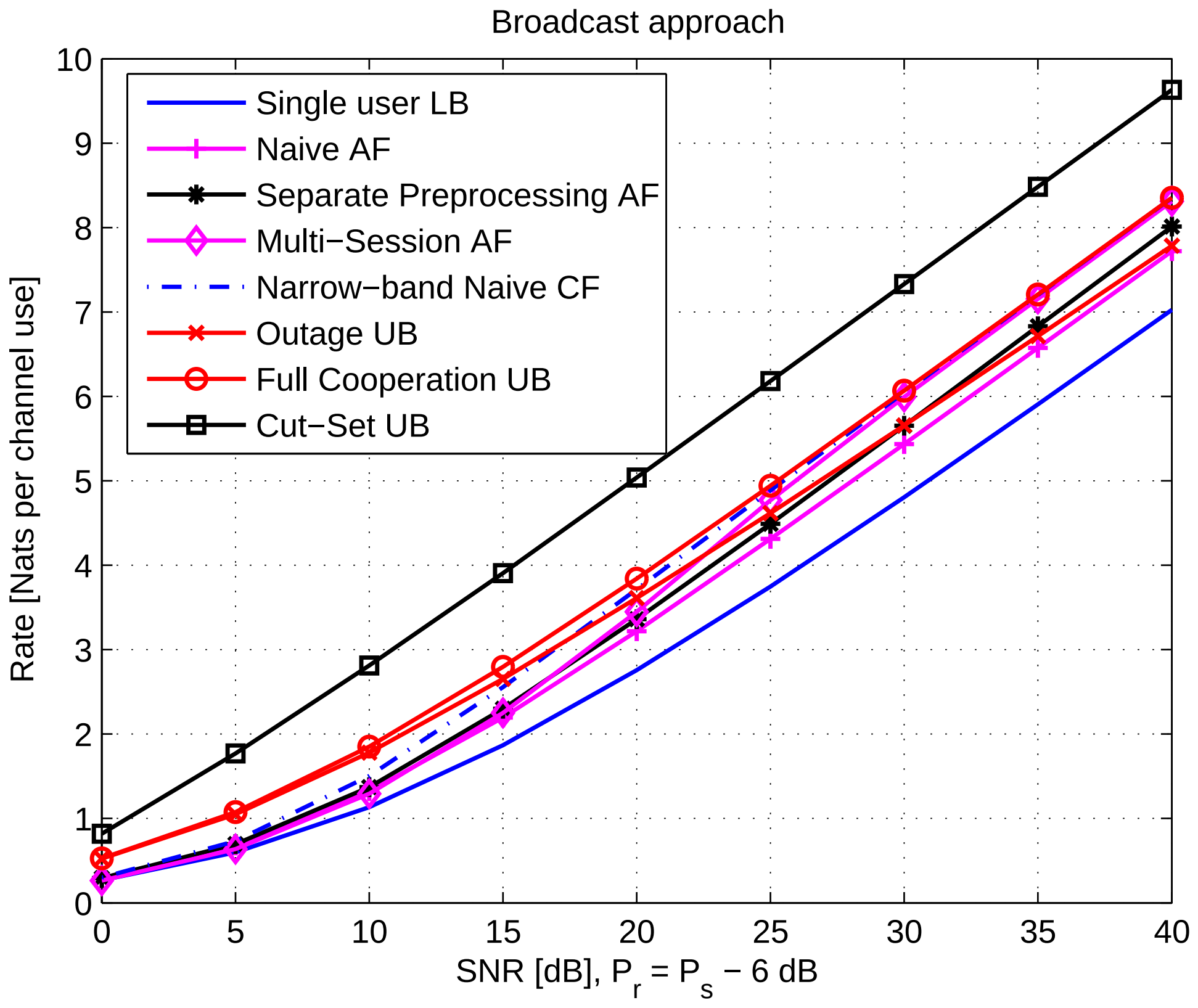

Broadcast approach
Single user LB
Naive AF
Separate Preprocessing AF
Multi−Session AF
Narrow−band Naive CF
Outage UB
Full Cooperation UB
Cut−Set UB
Rate [Nats per channel use]
SNR [dB], P_r = P_s − 6 dB
0
1
2
3
4
5
6
7
8
9
10
0
5
10
15
20
25
30
35
40

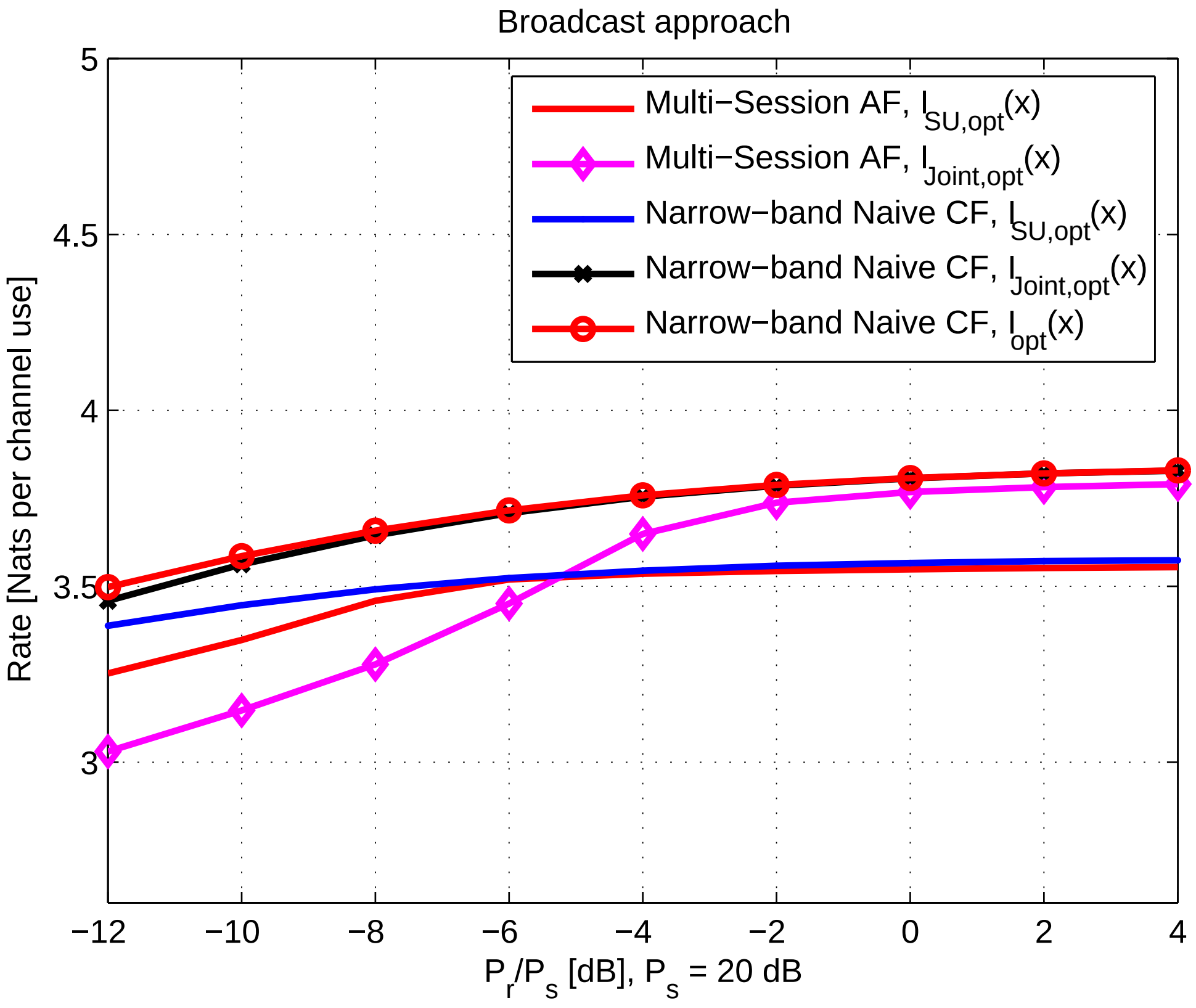
Broadcast approach
Multi−Session AF, I_SU,opt(x)
Multi−Session AF, I_Joint,opt(x)
Narrow−band Naive CF, I_SU,opt(x)
Narrow−band Naive CF, I_Joint,opt(x)
Narrow−band Naive CF, I_opt(x)
Rate [Nats per channel use]
5
4.5
4
3.5
3
−12
−10
−8
−6
−4
−2
0
2
4
P_r/P_s [dB], P_s = 20 dB

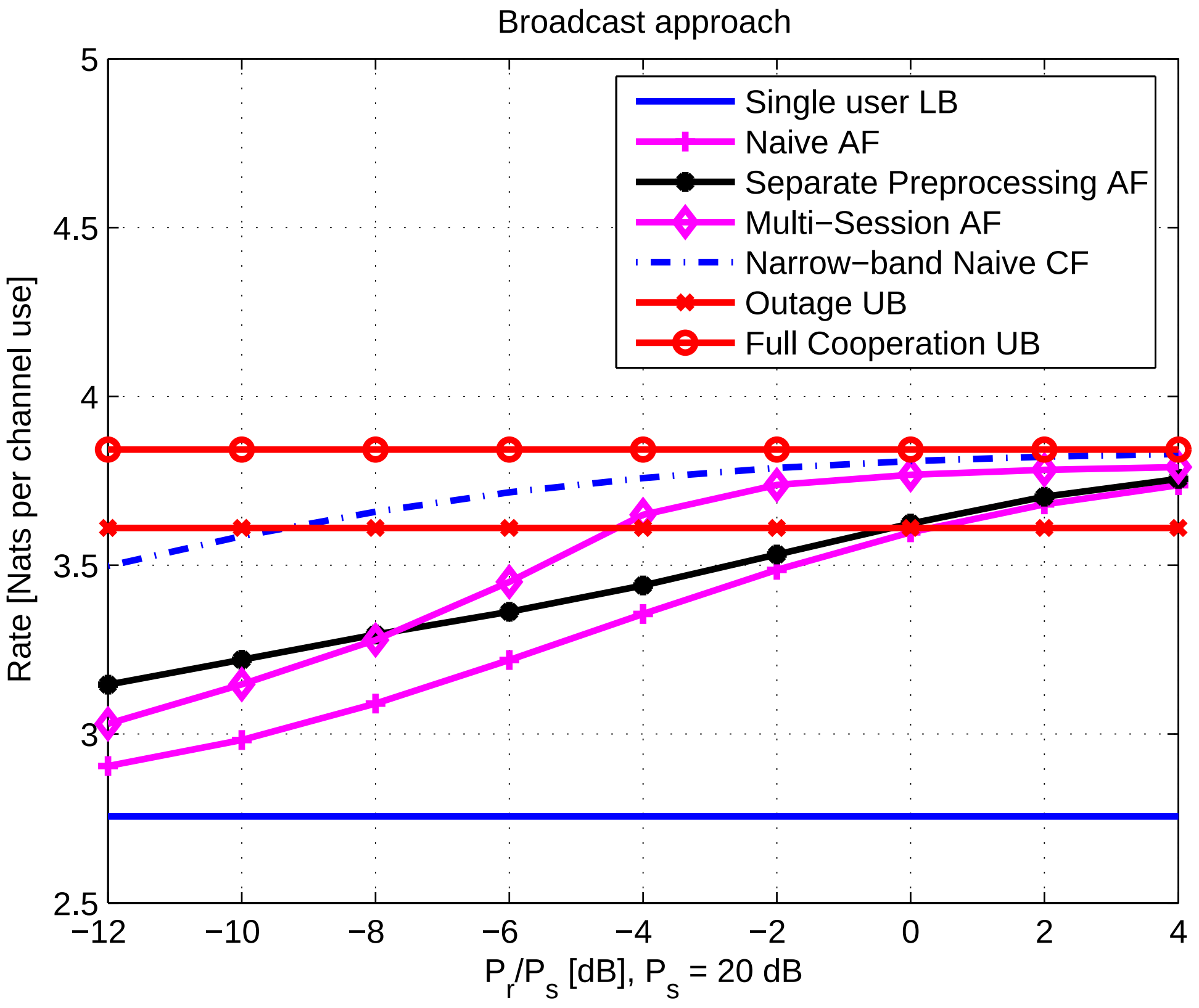

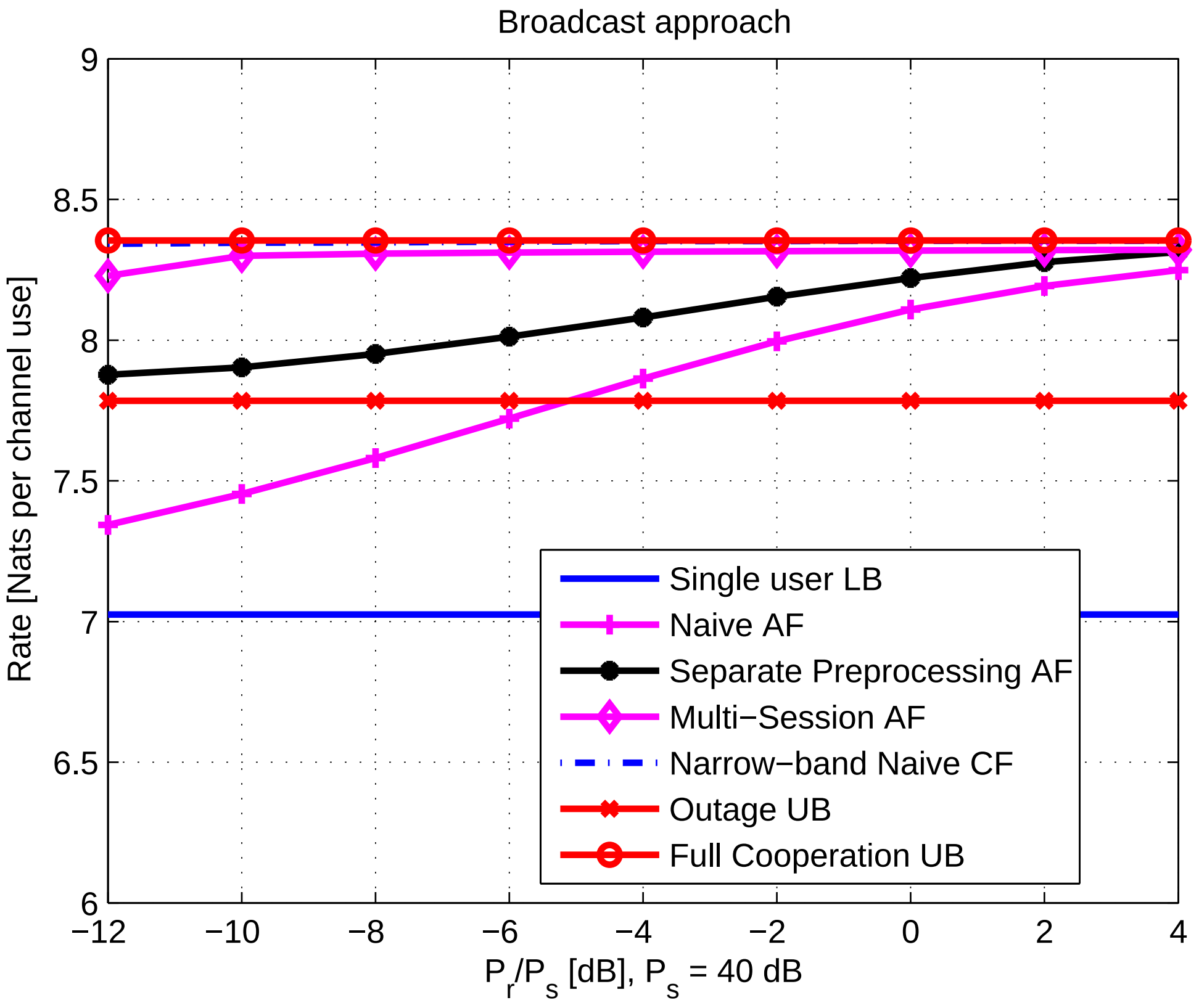

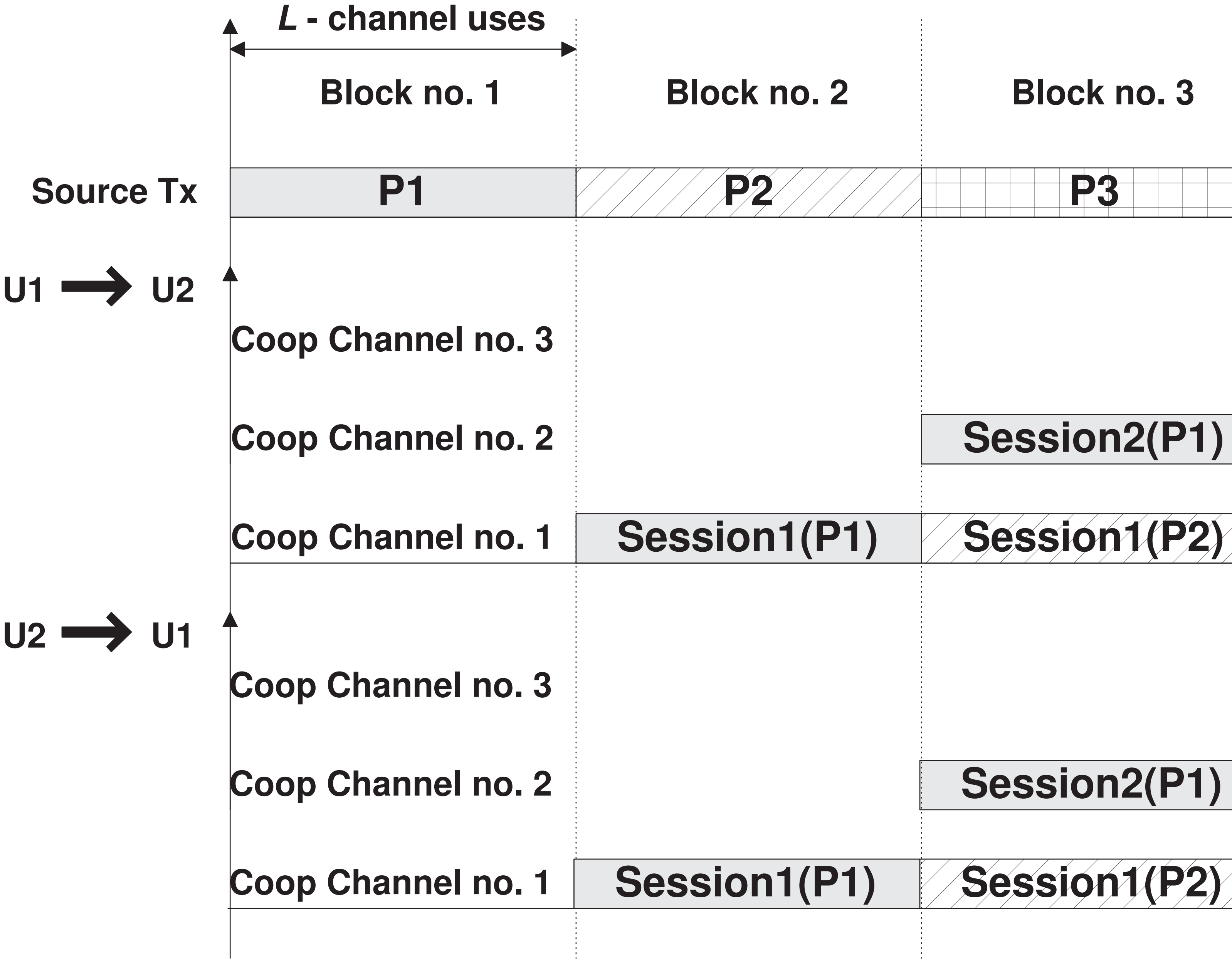

L - channel uses
Block no. 1
Block no. 2
Block no. 3
Source Tx
P1
P2
P3
U1 → U2
Coop Channel no. 3
Coop Channel no. 2
Session2(P1)
Coop Channel no. 1
Session1(P1)
Session1(P2)
U2 → U1
Coop Channel no. 3
Coop Channel no. 2
Session2(P1)
Coop Channel no. 1
Session1(P1)
Session1(P2)

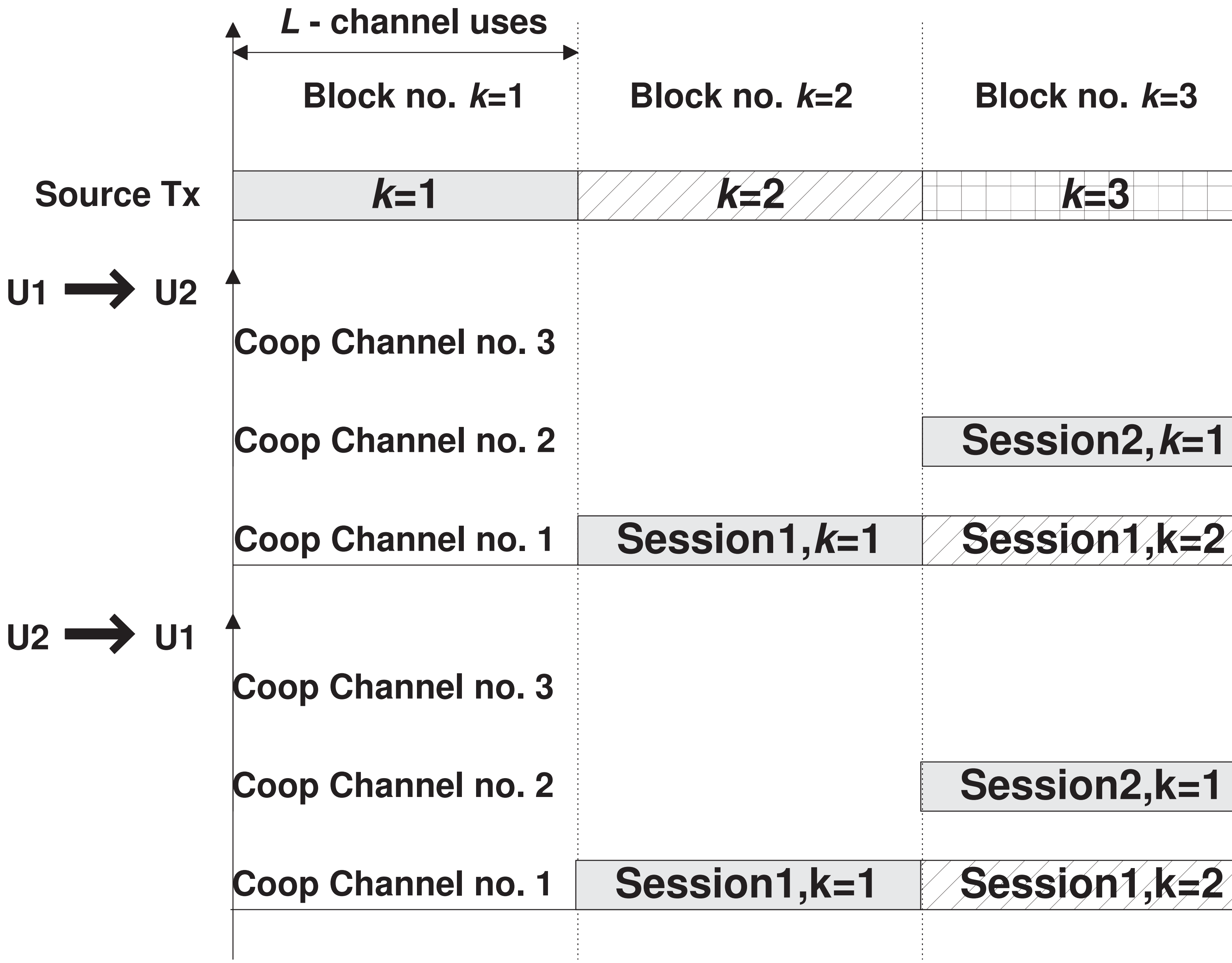
L - channel uses
Block no. k=1
Block no. k=2
Block no. k=3
Source Tx
k=1
k=2
k=3
U1 → U2
Coop Channel no. 3
Coop Channel no. 2
Session2, k=1
Coop Channel no. 1
Session1, k=1
Session1,k=2
U2 → U1
Coop Channel no. 3
Coop Channel no. 2
Session2,k=1
Coop Channel no. 1
Session1,k=1
Session1,k=2

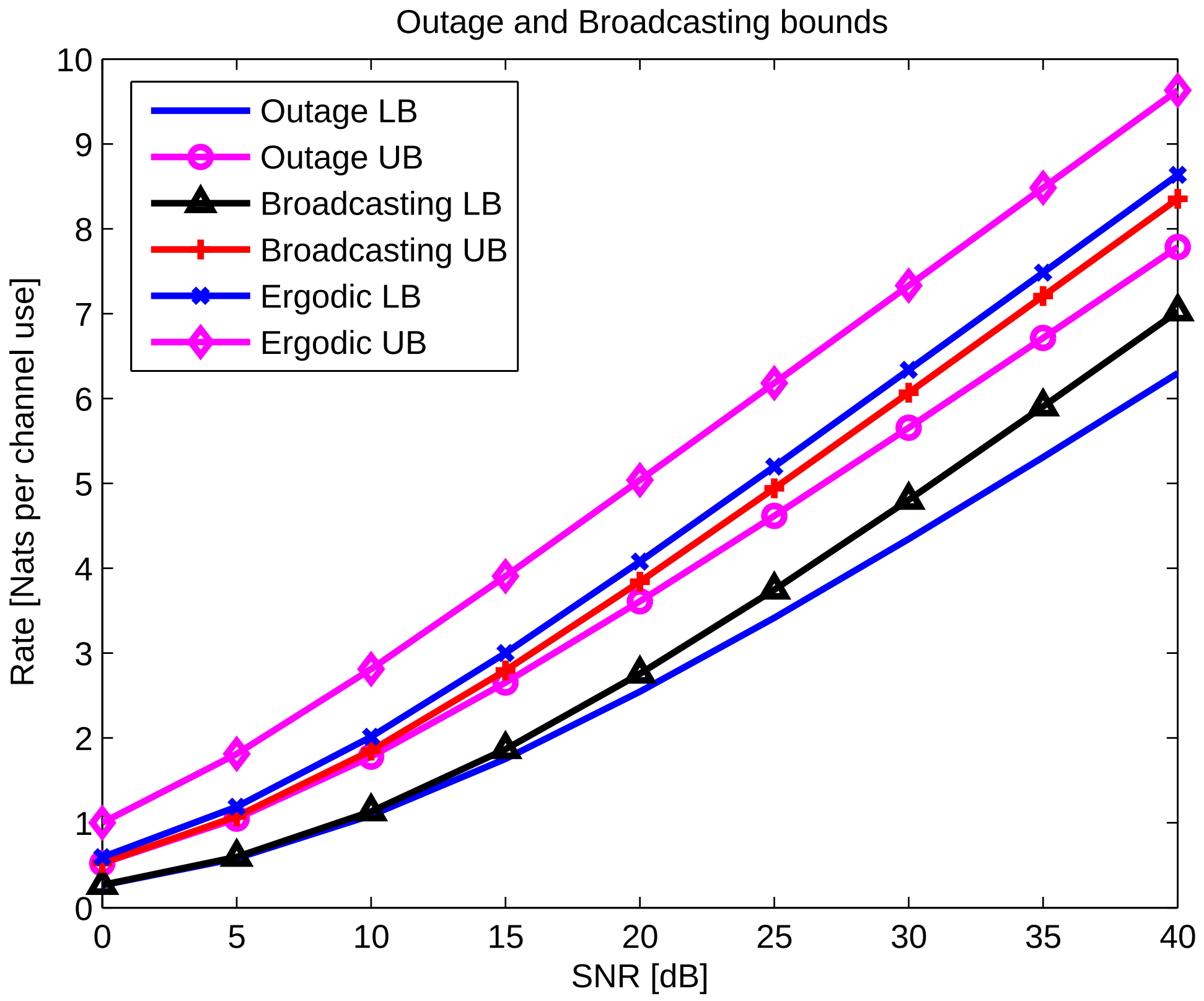

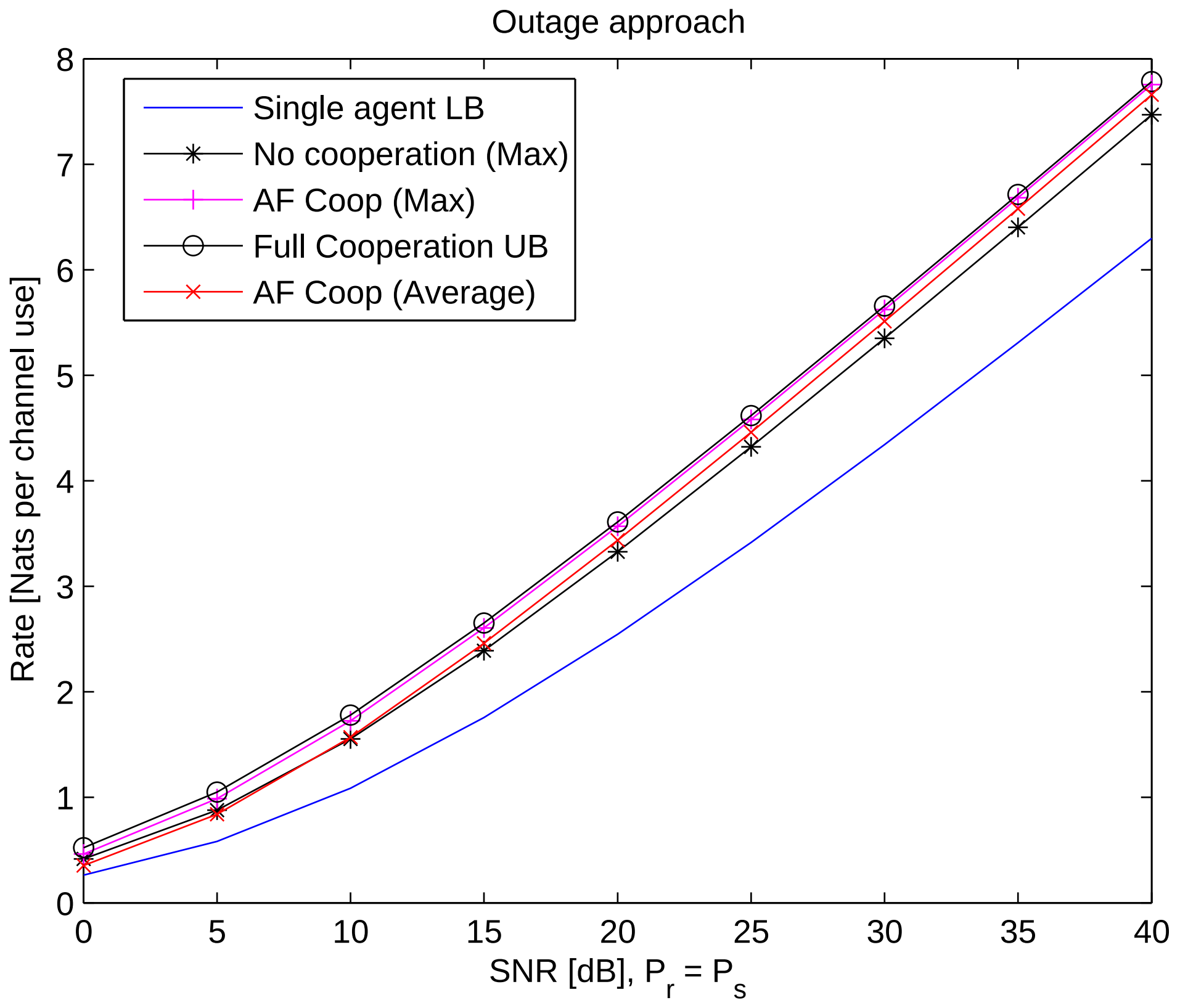

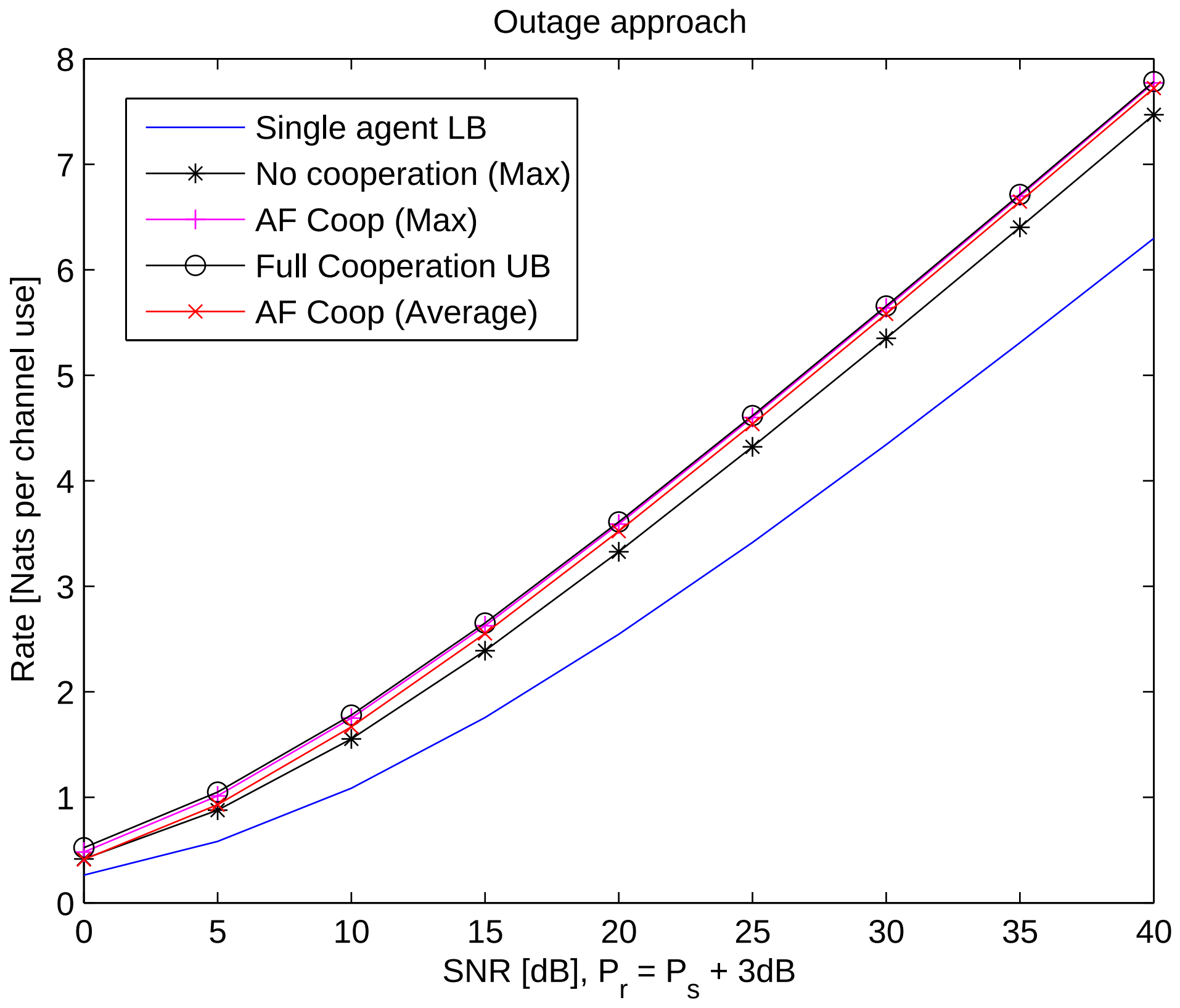

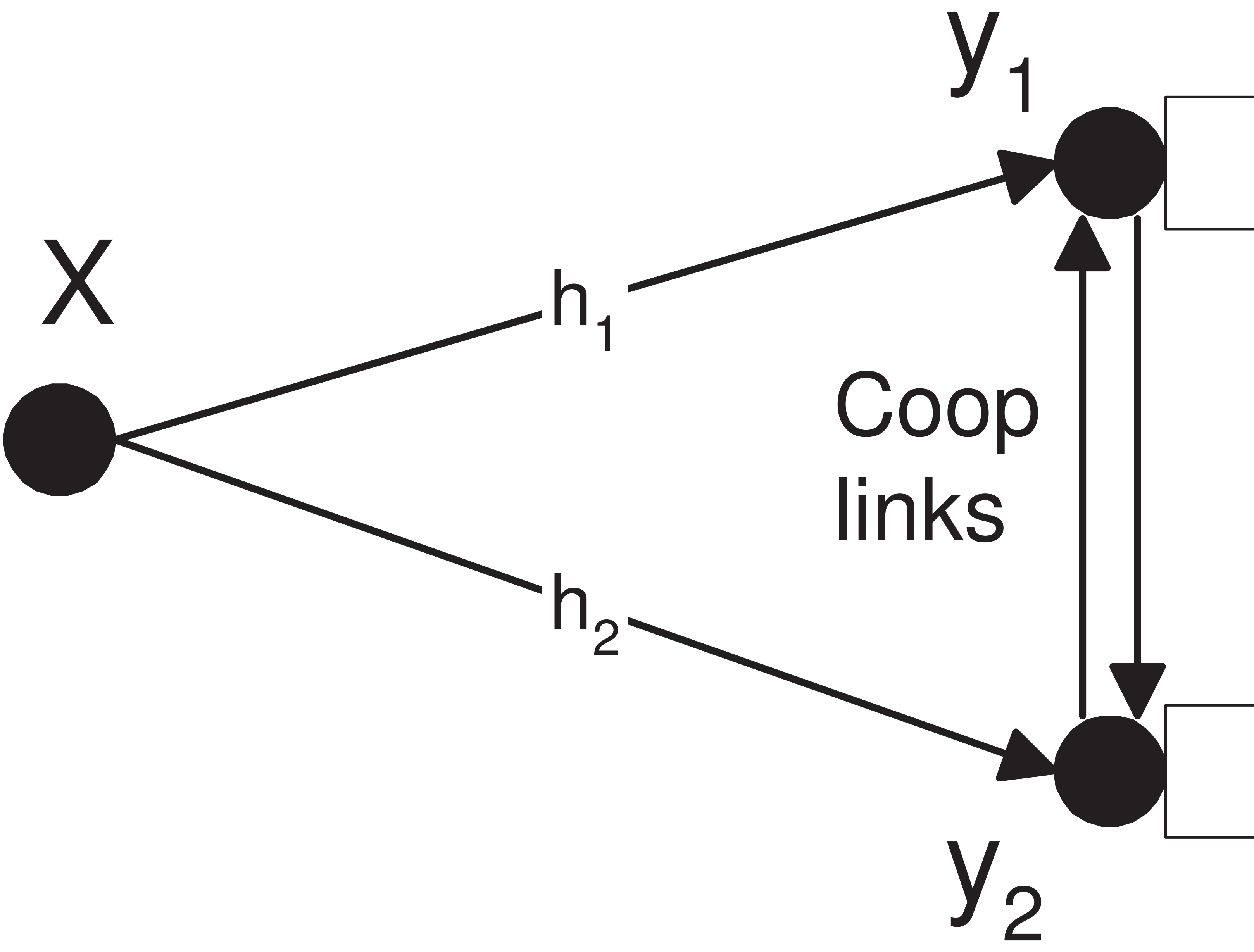

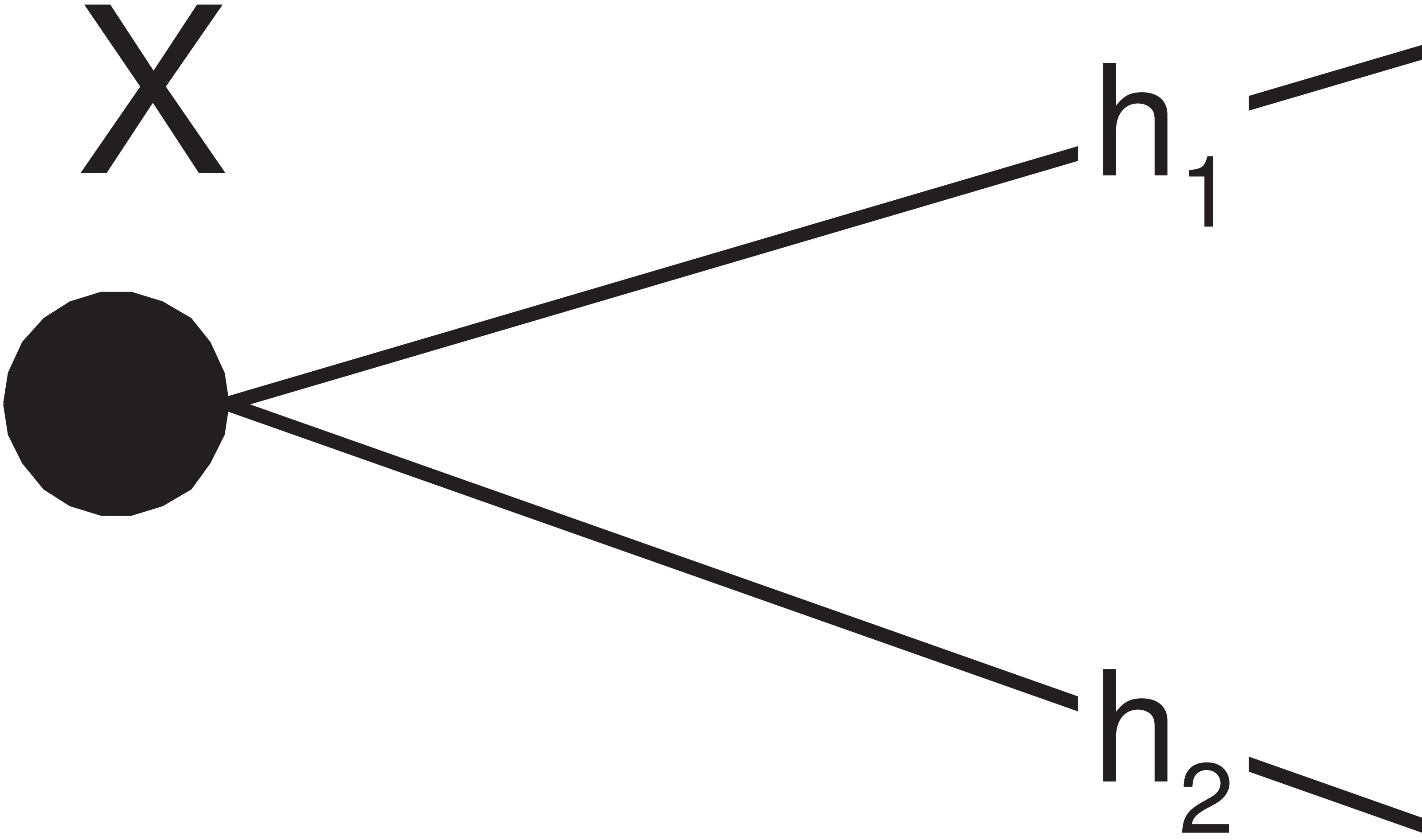

X
h1
h2